\documentclass[a4paper,12pt,twoside,reqno]{amsart}

\usepackage[T1]{fontenc}
\usepackage[utf8]{inputenc}
\usepackage[english]{babel}

\usepackage[osf]{mathpazo}
\usepackage{eucal}
\usepackage{mathrsfs}
\usepackage{microtype}

\usepackage{amsmath,amssymb,amsthm}
\usepackage{mathtools}
\usepackage{esint}
\usepackage{upgreek}

\usepackage{graphicx}
\usepackage{array}
\usepackage{enumitem}
\usepackage{xcolor}
\usepackage{setspace}
\usepackage{tikz}
\usepackage{tikz-cd}
\usepackage{hyperref}
\usepackage{xfrac}
\allowdisplaybreaks

\makeatletter
\@ifundefined{CMcal}{\DeclareMathAlphabet{\CMcal}{OMS}{cmsy}{m}{n}}{}
\makeatother

\theoremstyle{plain}
\newtheorem{theorem}{Theorem}[section]
\newtheorem{corollary}[theorem]{Corollary}
\newtheorem{proposition}[theorem]{Proposition}
\newtheorem{lemma}[theorem]{Lemma}

\theoremstyle{definition}
\newtheorem{definition}[theorem]{Definition}

\theoremstyle{remark}
\newtheorem{remark}[theorem]{Remark}
\newtheorem{example}[theorem]{Example}

\numberwithin{equation}{section}
\numberwithin{figure}{section}
\numberwithin{table}{section}

\newcommand{\R}{\mathbb{R}}
\newcommand{\N}{\mathbb{N}}
\newcommand{\C}{\mathbb{C}}

\newcommand{\Z}{\mathbb{Z}}

\newcommand{\s}[1]{\CMcal{#1}}

\newcommand{\bb}[1]{\mathscr{#1}}

\newcommand{\bra}[1]{\langle#1|}
\newcommand{\ket}[1]{|#1\rangle}

\newcommand{\ketbra}[2]{|#1\rangle\langle#2|}

\newcommand{\ii}{\mathrm{i}}

\DeclarePairedDelimiter{\norm}{\lVert}{\rVert}

\newcommand{\ie}{\textsl{i.\,e.\,}}

\hypersetup{
  colorlinks=true,
  linkcolor=purple,
  citecolor=teal,
  urlcolor=blue,
  filecolor=magenta,
  breaklinks=true,
  bookmarksopen=true,
  pdftitle={Fr\'echet structure over AF algebras I: the compact case},
  pdfauthor={Giuseppe De Nittis, Juli\'an G\'omez},
}

\begin{document}

\title[Fr\'echet structure over AF algebras I]{
Fr\'echet structure over AF algebras I: the compact case}

\author[G. De~Nittis]{Giuseppe De Nittis}

\author[J. G\'omez]{Juli\'an G\'omez}

\address[G. De~Nittis]{Facultad de Matem\'aticas \& Instituto de F\'isica,
  Pontificia Universidad Cat\'olica de Chile, Santiago, Chile.}
\email{gidenittis@uc.cl}

\address[J. G\'omez]{Departamento de Matem\'aticas,
  Universidad de los Andes, Bogot\'a, Colombia \&
  Facultad de Matem\'aticas,
  Pontificia Universidad Cat\'olica de Chile, Santiago, Chile.}
\email{jd.calderon@uniandes.edu.co}

\date{\today}
% [MSC] one primary and three secondary classes proposed by the editorial pass; to be checked
% MSC2020 list before submission (see the companion revision note, item [E3]).

\begin{abstract}
This is the first of two papers devoted to spatial Fréchet structures on AF algebras and their derivations. Here, we study the simplest example of such a structure: the Fréchet algebra of almost-local compact operators associated with a fixed orthonormal basis. We revisit the characterization of its continuous derivations in terms of infinite matrices and their operator-theoretic implementation on rapidly decreasing vectors. Reformulating this description in the language of 0-chains, borrowed from the theory of quantum many-body systems, we establish a sufficient condition for continuity in terms of spatial growth. Finally, we construct a Fréchet-continuous symmetric derivation that does not generate a one-parameter group of automorphisms of this algebra that is strongly continuous with respect to its Fréchet topology.
\end{abstract}

\maketitle
\tableofcontents

\section{Introduction}\label{sec:Intr0}

With the emergence of the study of $C^*$-algebras in the 40's, strongly motivated by their applications in statistical mechanics, the study of derivations on these algebras has proven to be of fundamental importance. Sakai~\cite{sakaiBounded} studied globally defined derivations on $C^*$-algebras, famously establishing that they are automatically continuous. Moreover, Kadison and Ringrose~\cite{kadison} (see also~\cite{Sakai}) showed that, when a $C^*$-algebra is faithfully represented on a Hilbert space $\s{H}$, every symmetric derivation $\delta$ can be implemented by a bounded self-adjoint operator $H_\delta$ in the double commutant of the represented algebra, in the sense that
$$
\delta(A) \; = \; \ii [H_\delta, A] \; .
$$

 \smallskip

 The case of unbounded derivations has also been extensively studied. In particular, Bratteli and Robinson~\cite{articleBratteli} established conditions under which such derivations admit an implementation by possibly unbounded self-adjoint operators and generate strongly continuous one-parameter groups of automorphisms. However, verifying these conditions, which may involve the existence of a dense set of analytic elements or suitable resolvent properties, can be difficult in practice. 

 \smallskip

 This difficulty is particularly relevant in statistical mechanics, where unbounded derivations arise naturally in the study of quantum many-body systems. In this setting, derivations are typically described in terms of local ``interactions'' which, in the so-called thermodynamic limit, give rise to derivations that cannot be defined on the entire algebra of observables. Furthermore, from a physical perspective, an important question is whether these local interactions give rise to a well-defined global dynamics on the algebra of observables in the thermodynamic limit.

 \smallskip

 A fundamental tool for addressing this question is provided by the so-called Lieb--Robinson bounds, first established by Lieb and Robinson in 1972~\cite{LiebRobinsonOriginal}. These bounds quantify the spatial propagation of the local dynamics through the system and, under suitable conditions on the interactions, play a central role in establishing the existence of infinite-volume dynamics (for a comprehensive treatment of Lieb--Robinson bounds, see~\cite{Nachtergaele-Sims-Young-Quasilocality-bounds}). Understanding the dynamics generated by local interactions and their localization properties remains an active area of research in modern mathematical physics (see, for instance,~\cite{Teufel-Generated-Spatially-Growing}).

 \smallskip

 A key ingredient in the study of these systems is played by a subalgebra that encodes how rapidly an element can be approximated by local observables with respect to the underlying spatial geometry. By equipping this space with a family of seminorms that quantify this approximation property, one obtains a Fréchet algebra of \emph{almost-local observables}. These objects were originally introduced in algebraic quantum field theory~\cite{araki1967collision} and subsequently adapted to the lattice setting in~\cite{bachmann2016lieb, ogata2021Cohom, Kapustin-Sopenko-Local-Noether}. Within the framework of quantum many-body systems, this algebraic structure provides a natural setting in which to study local interactions, their associated dynamics, and the corresponding localization properties. In particular, the relationship between Lieb--Robinson bounds and continuity with respect to the Fréchet topology of almost-local observables has been studied in~\cite{JulianLiebRobinson}.

 \smallskip

 From a more general mathematical perspective, one of the important features of quantum spin systems is the presence of an underlying separable metric space $(X,d)$ and a family of finite-dimensional local algebras whose union is dense in the global $C^*$-algebra, which is an AF algebra. In particular, to each site $v \in X$, one can associate an increasing family of finite-dimensional algebras $\{\bb{A}_v(n)\}_{n \in \N}$, where the geometry of the space $X$ plays a role in determining how the families associated with different sites and levels fit together. A similar spatial organization can also be found in certain families of graph $C^*$-algebras, particularly those associated with acyclic graphs (for a comprehensive treatment of graph $C^*$- algebras see~\cite{Raeburn}). This common structure suggests a more general construction of Fréchet subalgebras, whose seminorms quantify how rapidly an element can be approximated by these finite-dimensional algebras, taking into account both the geometry of the underlying metric space and the different levels of the corresponding inductive structures.

 \smallskip

 These observations motivate a broader problem, which we address in a series of two papers, the present one being the first. Our aim is to identify sufficient conditions under which spatially defined derivations are well-defined and continuous with respect to the underlying Fréchet topology, and to determine when they generate strongly continuous one-parameter groups of automorphisms on the corresponding $C^*$-algebra. The study of these more general spatial structures, together with sufficient conditions for dynamical generation, will be addressed in the second part of this work.

 \smallskip

In the present paper, we restrict our attention to the algebra of compact operators, which provides a simpler example of an AF algebra admitting such a spatial structure, with associated metric space $X = \Z$. From a physical perspective, this setting is also closely related to the single-particle tight-binding approximation, providing a simplified lattice model in which the role of the underlying geometry remains explicit. The Fréchet algebra of compact operators arising from this spatial structure, $\bb{K}_\infty$, was studied in~\cite{Klimek}, where the authors also provided a characterization of its continuous derivations by associating to each such derivation a family of coefficients $\{\beta_{n,l}\}_{n, l \in \Z}$ satisfying certain growth conditions. These coefficients, in turn, determine a formal infinite matrix $\mathsf{B}^\delta = (\mathsf{B}^\delta_{k,l})_{k,l \in \Z}$. We revisit this characterization and provide a proof in Theorem~\ref{Theorem: Continuous derivations}.

 \smallskip

We then describe continuous derivations in terms of operators acting on the Fréchet space $\s{H}_\infty$ of rapidly decreasing vectors. More precisely, we show that every Fréchet-continuous derivation on $\bb{K}_\infty$ can be implemented by an operator that leaves $\s{H}_\infty$ invariant and is continuous with respect to its Fréchet topology. The same properties hold for the restriction of its adjoint operator to $\s{H}_\infty$. Conversely, operators satisfying these conditions give rise to Fréchet-continuous derivations on $\bb{K}_\infty$ (Proposition~\ref{Proposition: Domain Bdelta}). This operator-theoretic description will be useful later when studying the generation of dynamics.

 \smallskip

We then reformulate the preceding results using the language of $0$-chains (see Definition~\ref{Def: 0-Chains}),
$$
\Phi : \Z \longrightarrow \bb{K}_\infty \; ,
$$
a formalism that has also been introduced in the study of quantum spin systems~\cite{Kapustin-Sopenko-Local-Noether}. This description allows us to study derivations through the spatial growth of the individual terms $\Phi(k)$. In particular, using the characterization of~\cite{Klimek}, we show that every Fréchet-continuous symmetric derivation on $\bb{K}_\infty$ can be represented by a $0$-chain whose spatial norms grow at most polynomially in $|k|$. More precisely, one can choose a representative satisfying estimates of the form
$$ |||\Phi(k)|||_{k, \alpha} \; \leqslant \; C_\alpha (1 + |k|)^{\tau(\alpha)}, \qquad k \in \Z,\quad \alpha \in\N_0,
$$
for suitable constants $C_\alpha > 0$ and exponents $\tau(\alpha) \geq 0$ (see Proposition~\ref{Prop: interaction and derivation} and Eq.~\eqref{eq: From derivation to interaction spatial estimate}).

 \smallskip

This observation suggests that controlling the spatial growth of $0$-chains may provide sufficient conditions for continuity, even when the corresponding spatial norms are not uniformly bounded. One of the main results of this work establishes such a condition for the compact-operator algebra $\bb{K}_\infty$. More precisely, we prove the following.

\begin{theorem}[Sufficient growth conditions for continuity]\label{Theorem 1 intro}
Let $\Phi$ be a $0$-chain. Suppose that, for every $\alpha \in \N_0$, there exist constants $C_\alpha > 0$ and $s_\alpha \in \R_{\geq 0}$ such that
$$ |||\Phi(k)|||_{k, \alpha} \; \leqslant \; C_\alpha (1 + |k|)^{s_\alpha}, \qquad k \in \Z, $$
and that
$$ \limsup_{\alpha \to \infty}(\alpha - s_\alpha) \; = \; \infty \; .$$
Then the corresponding formal derivation 
$$\delta_\Phi (A) \; := \; \ii \sum_{k \in \Z}[\Phi(k), A]$$ 
is well-defined and Fréchet-continuous on $\bb{K}_\infty$.
\end{theorem}

Although this result does not provide a full characterization of continuous derivations on $\bb{K}_\infty$, it gives a sufficient condition for continuity formulated directly in terms of the spatial growth of $0$-chains (Proposition~\ref{Prop: Growth-Condition 0 chain}). More importantly, this formulation provides a starting point for studying continuous derivations on the more general spatial Fréchet algebras that we intend to consider in the future work. Related questions have also been studied in the context of quantum spin systems. In particular, in~\cite{Teufel-Generated-Spatially-Growing}, the authors consider spatially growing interactions whose local terms are allowed to grow at most linearly, and establish conditions under which the corresponding derivations generate well-defined dynamics.

 \smallskip

This brings us to another question in our setting: whether the growth conditions introduced above, which guarantee the continuity of the associated derivation, are also sufficient to ensure that it generates a strongly continuous one-parameter group of automorphisms. The general problem of integrating derivations into semigroups or strongly continuous groups of operators on locally convex and Fréchet spaces was already addressed by Yosida and Phillips (see, for instance,~\cite[Chapter IX]{Yosida}). However, the abstract functional-analytic resolvent conditions required by their results are often difficult to verify in practice, particularly for derivations arising from local interactions or $0$-chains.

 \smallskip

\noindent In the present work, we show that the answer to this question is negative, even in the case of compact operators.

\begin{theorem}[Continuity does not imply dynamical generation]\label{Theorem 2 intro}
There exists a $0$-chain $\Phi$ satisfying the growth conditions stated above such that its associated derivation $\delta_\Phi$ is Fréchet-continuous on $\bb{K}_\infty$, but fails to generate a one-parameter group of automorphisms of $\bb{K}_\infty$ that is strongly continuous with respect to the Fr\'echet topology.
\end{theorem}

An important step in establishing this result is the relation between dynamical generation and the essential self-adjointness of the operators associated with symmetric derivations. In particular, we show that, whenever a Fréchet-continuous symmetric derivation generates a strongly continuous one-parameter group of automorphisms, its associated symmetric operator on $\s{H}_\infty$ must be essentially self-adjoint (Corollary~\ref{Corollary: Necessary condition essential self-adjointness}). We then construct an explicit example, based on Jacobi matrices, whose associated symmetric operator fails to be essentially self-adjoint. This provides the counterexample stated above.

 \smallskip

Our results thus show that controlling the spatial growth of a $0$-chain, while sufficient to guarantee continuity of the corresponding derivation, does not necessarily ensure dynamical generation. They also motivate the search for additional conditions under which such derivations generate well-defined dynamics, a problem we will address in the second paper of this series.

 \smallskip

{\bf Structure of the paper.}
In Section~\ref{Section 1}, we construct the Fr\'echet $*$-algebra $\bb{K}_\infty$ of almost-local compact operators, starting from a fixed orthonormal basis $\Psi$. We present two equivalent descriptions of this algebra: one based on rapid approximation by localized operators and the other in terms of matrix coefficients. In Section~\ref{Section: Frechet continuous derivations}, we revisit the characterization of Fr\'echet-continuous derivations on $\bb{K}_\infty$ and provide an operator-theoretic interpretation of this characterization. Section~\ref{Section: 0-chains} introduces the formalism of $0$-chains and establishes sufficient conditions on their spatial growth for the continuity of the associated derivations, proving Theorem~\ref{Theorem 1 intro}. Finally, in Section~\ref{Section: Non-integrable derivations}, we construct a Fr\'echet-continuous symmetric derivation that fails to generate a Fr\'echet-strongly continuous one-parameter group of automorphisms, proving Theorem~\ref{Theorem 1 intro}.

 \smallskip
 
 \noindent
{\bf Acknowledgements.}
The authors gratefully acknowledge the support provided by the grant \emph{Fondecyt Regular - 1230032}. J.G. also thanks the Vice Presidency of Research and Creation at Universidad de los Andes for its financial support through project No. INV-2025-220-3480, as well as the Faculty of Sciences of Universidad de los Andes through project No. INV-2026-230-3641. 
%The authors would also like to thank Sven Bachmann, Rahul Hingorani, Bruno Nachtergaele, Stefan Teufel for numerous stimulating and insightful discussions related to this work.

 \smallskip
 
 \noindent
{\bf Conflict of interest statement.}
 The authors certify that they have no affiliations with or involvement in any organization or entity with any financial interest (such as honoraria; educational grants; participation in speakers' bureaus; membership, employment, consultancies, stock ownership, or other equity interest; and expert testimony or patent-licensing arrangements), or non--financial interest (such as personal or professional relationships, affiliations, knowledge or beliefs) in the subject matter or materials discussed in this manuscript.

\smallskip

\noindent
{\bf Data availability statement.}
No datasets were generated or analyzed during the current study. The scripts used to perform the symbolic and numerical verifications supporting the results of this work are available from the authors upon reasonable request.

\smallskip
\noindent
{\bf Use of artificial intelligence statement.}
During the preparation of this work, the authors used Claude \emph{Opus 5} (Anthropic) to improve the linguistic quality, grammatical accuracy, and readability of the manuscript. The tool was also employed to cross-check the algebraic consistency and overall correctness of the calculations presented in the various theorems. All AI-generated outputs were subsequently thoroughly reviewed, independently verified, and edited by the authors, who take full responsibility for the mathematical accuracy, academic integrity, and final scientific content of the published article.
%as well as to support and streamline the bibliographic search process. 

\newpage

\section{The Fréchet algebra of compact operators}\label{Section 1}

In this section, we study the spatial Fréchet structure associated with the $C^*$-algebra of compact operators $\bb{K} \equiv \bb{K}(\s{H})$ on a separable Hilbert space $\s{H}$. Taking $\Z$ as the underlying discrete metric space, we construct the Fréchet $*$-algebra $\bb{K}_\infty$ of \emph{almost-local} compact operators and examine its basic properties. This construction establishes the setting for our subsequent study of continuous derivations and their description in terms of $0$-chains.

\subsection{Fréchet structure}\label{Subsection: Frechet structure}

In order to define local and almost local structures on the algebra of compact operators, we start by fixing an orthonormal basis $\Psi = \{\psi_n\}_{n \in \Z}$ for the Hilbert space $\s{H}$. Let $\N_0:=\N\cup\{0\}$. Given $j \in \Z$ and $r \in \N_0$, let $V_{j,r}$ be the finite dimensional subspace spanned by $\{\psi_{j-r}, \ldots, \psi_j, \ldots, \psi_{j + r}\}$, and let $\bb{K}_{j,r}$ be the subalgebra of operators acting on $V_{j,r}$, with a trivial action on $V_{j,r}^\bot$. There is a natural inclusion $\bb{K}_{j,r} \xhookrightarrow{} \bb{K}_{j, r + 1}$, which in matrix notation is given as $A \mapsto {\rm Diag }(0, A, 0)$. These operators can be understood as finitely supported compact operators with respect to the basis $\Psi$. Let 
$$\bb{K}_{j,{\rm loc}} \;:= \;\bigcup_{r \in \N_0} \bb{K}_{j, r}$$
be the algebra of these finitely supported operators initially centered in $j \in \Z$. Notice that this collection of subalgebras $\{\bb{K}_{j, r}\}_{r\in\N_0}$ defines the natural AF-structure of the algebra of compact operators $\bb{K}$. Furthermore, this dense local *-subalgebra does not depend on the choice of the center $j \in \Z$. In fact, given $ j \neq k$ and $r \in \N_0$, one has that $\bb{K}_{j,r} \subseteq \bb{K}_{k, r + |j - k|}$. Consequently, we simplify the notation and denote this algebra simply as $\bb{K}_{\rm loc}$.

 \smallskip

For each $j \in \Z$ and $r \in \N_0$, let $P_{j,r}$ be the orthogonal projection onto the closed subspace $V_{j,r}$. Let $U_\Psi: \s{H} \rightarrow \s{H}$ be the right-shift operator with respect to the basis $\Psi$, that is, $U_\Psi$ acts as $\psi_n \mapsto \psi_{n + 1}.$ Hence, $U^{-1}_\Psi$ corresponds to the left-shift operator. Observe that for any $j \in \Z$ and $r \in \N_0$, one has $U^{j}_\Psi P_{0,r} U^{-j}_\Psi = P_{j,r}$.
Using these orthogonal projections, we define the family of functions $p_j: \N_0 \times \bb{K} \rightarrow \R_{\geq 0}$ as
\begin{equation*}
	p_j(r, A) \; := \; ||A - P_{j,r} A P_{j, r}||\;.
\end{equation*}
Notice that $p_j(r, A) \leqslant 2||A||$. Similarly, we define the family of functions $f_j: \N_0 \times \bb{K} \rightarrow \R_{\geqslant 0 }$ as
\begin{equation*}
	f_j(r, A) \; := \; \inf_{B \in \bb{K}_{j, r}} ||A - B||\;.
\end{equation*}
One easily checks that for any $j,k \in \Z$, $f_j(r + |j - k|, A) \leqslant f_k(r, A)$. Moreover, since $\bb{K}_{j,r}$ is a finite dimensional vector space, the infimum in the definition of the function $f_j$ is attained for any $A \in \bb{K}$ and $r \in \N_0$. Indeed, any minimizing sequence $\{C_n\}_{n \in \N_0}$ is bounded, and since $\bb{K}_{j,r}$ is finite-dimensional, it admits a convergent subsequence $\{C_{n_k}\}_{k \in \N_0}$. Hence,
$$f_j(r, A) \; = \; \lim_{k \to \infty} ||A - C_{n_k}|| \; = \; ||A - \lim_{k \to \infty} C_{n_k}||.$$ 

\noindent In the next result we show that these two functions, $f_j(r, A)$ and $p_j(r, A)$, are equivalent up to multiplicative constants. Their role is to make precise a notion of locality with respect to the basis $\Psi$. 
\begin{lemma}\label{lemma: Relation f and p}
	For any $A \in \bb{K}$, $j \in \Z$ and $r\in \N_0$ one has that
	$$f_j(r,A)\; \leqslant\; p_j(r,A) \;\leqslant\; 2 f_j(r,A)\;.$$
\end{lemma}
\begin{proof}
	The first inequality follows easily by observing that $P_{j,r} A P_{j,r}$ belongs to $\bb{K}_{j,r}$. Now, take an element $A|_r \in \bb{K}_{j,r}$ attaining the infimum in the definition of $f_j(r,A)$. Since for any element $B \in \bb{K}_{j,r}$, one clearly has $P_{j,r} B P_{j,r} = B$, it follows that 
	$$
	\begin{aligned}
	p_j(r,A) \; &= \; ||A - P_{j, r} A P_{j, r}|| \; \leqslant \; ||A - A|_r|| + ||P_{j, r}(A|_r -  A) P_{j, r}||\\
	& \leqslant\; 2 ||A - A|_r|| \;=\; 2f_j(r,A) \; . 
	\end{aligned}
	$$ 
	This concludes the proof.
\end{proof}

 \smallskip

Using these families, we construct the $*$-subalgebra of rapidly approximate compact operators with respect to $\Psi$. We will refer to these as \emph{almost-local} compact operators. To do that, for $j \in \Z$ and $\alpha \in \N_0$, we define the norm
\begin{equation*}
|||A|||_{j,\alpha} \;:=\; \|A\| + \sup_{r \in \mathbb{N}_0} (1 + r)^\alpha f_j(r,A)\;.
\end{equation*}
Notice that for any $A \in \bb{K}_{\rm{loc}}$ and any $\alpha \in \N_0$, one has $|||A|||_{j,\alpha} < \infty$. Furthermore, observe that the point $j \in \Z$ does not play an essential role in this notion of localization. In fact, denoting $R = |j - k|$, one has that 
\begin{equation}\label{eq: equivalence norms f}
	\frac{1}{(1 + R)^\alpha} |||A|||_{j, \alpha}\; \leqslant\; |||A|||_{k, \alpha} \; \leqslant \; (1 + R)^\alpha |||A|||_{j, \alpha} \; .
\end{equation}
Similarly, we could also use the functions $p_j$ in the definition of these norms. Since they differ only by a constant factor of $2$, we will work with both interchangeably. 

 \smallskip

 We can now use these approximation norms to give a precise definition of almost-local compact operators. Recall the definition of the space of rapidly decreasing sequences $\s{S}(\N_0)$ in Appendix \ref{app:Rap_seq}.

\begin{definition}\label{def: rapidly approximate algebra}
	The space of almost-local compact operators is defined by
	$$\bb{K}_\infty \; := \; \{A \in \bb{K} \;|\; r \mapsto f_j(r, A) \in \s{S}(\N_0) \;\; \text{for some} \;\; j\in \Z\} \; .$$ 
\end{definition}

 \smallskip

Notice that, by Eq.~\eqref{eq: equivalence norms f}, if $r \mapsto f_j(r,A)$ belongs to $\s{S}(\N_0)$ for some $j \in \Z$, then the same holds for every $k \in \Z$. Thus, the definition of $\bb{K}_\infty$ is independent of the choice of the localization site. 

 \smallskip

The following proposition establishes that $\bb{K}_\infty$ naturally carries the structure of a Fréchet $*$-algebra.

\begin{proposition}\label{Prop: properties Frechet algebra compact case}
For each fixed $j \in \Z$, $\bb{K}_\infty$ corresponds to the Fréchet completion of $\bb{K}_{\rm loc}$ with respect to the family of norms $\{|||\cdot|||_{j, \alpha}\}_{\alpha\in\N_0}$. Moreover, $\bb{K}_\infty$ has the structure of a Fréchet $\ast$-algebra: for all $A,B \in \bb{K}_\infty$, $\alpha \in \N_0$,
  $$|||A\,B|||_{j, \alpha}\; \leqslant\; \frac{3}{2} \, |||A|||_{j, \alpha} \ |||B|||_{j, \alpha}\;, \qquad |||A^*|||_{j, \alpha} \; = \; |||A|||_{j, \alpha}\;.$$
\end{proposition}

\begin{proof}
We first observe that, for each $r \in \N_0$, the map
$$ A \mapsto f_j(r,A) $$
defines a seminorm on $\bb{K}$. Indeed, given $A, B \in \bb{K}$, choose $A|_r, B|_r \in \bb{K}_{j,r}$ such that
$$ f_j(r, A) \; = \; ||A - A|_r||, \qquad f_j(r, B) \; = \; ||B - B|_r||. $$
Then 
$$ f_j(r, A + B) \; \leqslant \; ||A + B - A|_r - B|_r|| \; \leqslant \; f_j(r, A) + f_j(r, B),$$
while homogeneity follows immediately.

\noindent We next establish the $\ast$-algebra properties. Let $A, B \in \bb{K}_\infty$, and choose $A|_r$ and $B|_r$ as above. Then
\begin{align*}
f_j(r, AB) \; &\leqslant \; ||AB - A|_r B|_r|| \\
&\leqslant \; ||AB- A|_r B + A|_r B - A|_r B|_r|| \\
&\leqslant \; f_j(r, A) \, ||B|| + ||A|_r|| \, f_j(r, B)\\
&\leqslant \; f_j(r, A) \, ||B|| + \left(||A|| + f_j(r, A) \right)f_j(r, B)\\
&\leqslant \; \frac{3}{2}||B|| \, f_j(r, A) + \frac{3}{2} ||A|| \, f_j(r, B) \; ,
\end{align*}
where the last inequality follows from $f_j(r,A) f_j(r,B) \leqslant \frac{1}{2}\big(||A|| f_j(r,B) + ||B|| f_j(r,A)\big)$. Consequently, for every $\alpha \in \N_0$,
$$ |||AB|||_{j,\alpha} \; \leqslant \; \frac{3}{2} |||A|||_{j,\alpha} \, |||B|||_{j,\alpha} \; .$$
Similarly, the involution satisfies
$$|||A^*|||_{j, \alpha} \; = \; |||A|||_{j, \alpha} \; .$$

\noindent We now turn to the completeness of $\bb{K}_\infty$. Let $\{A^{(n)}\}_{n \in \N}$ be a Cauchy sequence with respect to the family of Fréchet norms. Since
$$|||A^{(n)} - A^{(m)}|||_{j, \alpha}  \; = \; ||A^{(n)} - A^{(m)}|| + \sup_{r \in \N_0}(1 + r)^\alpha f_j(r, A^{(n)} - A^{(m)}), $$
the sequence is also Cauchy in the operator norm and therefore converges to some $A \in \bb{K}$.

\noindent For each $n \in \N$, define $g_n(r) := f_j(r, A^{(n)})$. By the reverse triangle inequality,
$$ \left| f_j(r, A^{(n)}) - f_j(r, A^{(m)}) \right| \; \leqslant \; f_j(r, A^{(n)} - A^{(m)}). $$
It follows that $\{g_n\}_{n \in \N}$ is a Cauchy sequence in $\s{S}(\N_0)$ with respect to the norms $||\cdot||_\alpha$ (see Eq.~\eqref{eq:seq_nor}). By Proposition~\ref{prop:compl}, there exists $h \in \s{S}(\N_0)$ such that $g_n$ converges to $h$ in this topology. In particular, $g_n(r) \to h(r)$ for each $r \in \N_0$.

\noindent On the other hand, setting $g(r) := f_j(r, A)$, we have
$$ \left| f_j(r, A^{(n)}) - f_j(r, A) \right| \; \leqslant \; f_j(r, A^{(n)} - A) \; \leqslant\; ||A^{(n)} - A||. $$
Hence, $g_n(r)$ also converges pointwise to $g(r)$, and therefore $h = g$. Since $h \in \s{S}(\N_0)$, we conclude that $A \in \bb{K}_\infty$.

\noindent It remains to show that $A^{(n)} \to A$ with respect to the Fréchet topology. Fix $\alpha \in \N_0$ and let $\varepsilon > 0$. Since $A^{(n)} \to A$ in the operator norm and $\{A^{(n)}\}_{n \in \N}$ is Cauchy with respect to $|||\cdot|||_{j, \alpha}$, there exists $N_\varepsilon \in \N_0$ such that, for all $n, m \geq N_\varepsilon$,
\begin{equation}\label{eq0: proof completeness}
||A - A^{(n)}|| \; < \; \frac{\varepsilon}{3} \; ,
\end{equation}
and
\begin{equation}\label{eq00: proof completeness}
|||A^{(n)} - A^{(m)}|||_{j,\alpha} \; < \; \frac{\varepsilon}{3} \; .
\end{equation}
Now, fix $n \geq N_\varepsilon$. For every $r \in \N_0$, the triangle inequality yields
\begin{equation*}
\begin{aligned}
(1 + r)^\alpha f_j(r, A - A^{(n)}) \; \leqslant \; &(1 + r)^\alpha f_j(r, A^{(n)} - A^{(m)})\\
&+ (1 + r)^\alpha f_j(r, A - A^{(m)}).
\end{aligned}
\end{equation*}
For every $m \geq N_\varepsilon$, by~\eqref{eq00: proof completeness}, the first term on the right-hand side is bounded by $\varepsilon/3$, uniformly in $r$. Moreover, since $A^{(m)} \to A$ in the operator norm, for each fixed $r \in \N_0$, we may choose $m \geq N_\varepsilon$ sufficiently large so that
$$ (1 + r)^\alpha f_j(r, A - A^{(m)}) \; \leqslant \; (1 + r)^\alpha ||A - A^{(m)}|| \; < \; \frac{\varepsilon}{3} \; .$$
With this choice, we obtain
$$(1 + r)^\alpha f_j(r, A - A^{(n)}) \; < \; \frac{2\varepsilon}{3} \; .$$
Since the bound holds for all $r \in \N_0$, taking the supremum yields
\begin{equation}\label{eq2: proof completeness}
\sup_{r \in \N_0}(1 + r)^\alpha f_j(r, A - A^{(n)}) \; \leqslant \; \frac{2\varepsilon}{3} \; .
\end{equation}
Combining Eqs.~\eqref{eq0: proof completeness} and~\eqref{eq2: proof completeness}, we conclude that, for every $n \geq N_\varepsilon$,
$$|||A - A^{(n)}|||_{j,\alpha} \; < \; \varepsilon \; .$$

\noindent Finally, we show that $\bb{K}_{\rm loc}$ is dense in $\bb{K}_\infty$ with respect to the Fréchet topology. Let $A \in \bb{K}_\infty$, and for each $n \in \N_0$, choose $A^{(n)} \in \bb{K}_{j,n}$ such that
$$ f_j(n, A) \; = \; ||A - A^{(n)}|| \; .$$
By the definition of the Fréchet norms,
$$ |||A - A^{(n)}|||_{j,\alpha} \; = \; f_j(n, A) + \sup_{r \geq 0}(1 + r)^\alpha f_j(r, A - A^{(n)}) \; .$$
For $r > n$, observe that $ f_j(r, A - A^{(n)}) = f_j(r, A)$. On the other hand, for $r \leq n$, we have
$$ (1 + r)^\alpha f_j(r, A - A^{(n)}) \; \leqslant \; (1 + n)^\alpha||A - A^{(n)}|| \; = \; (1 + n)^\alpha f_j(n, A) \; .$$
Hence,
$$|||A - A^{(n)}|||_{j,\alpha} \; \leqslant \; f_j(n, A) \, + \, \sup_{r \geq n}\,(1 + r)^\alpha f_j(r, A) \; ,$$
which tends to zero as $n \to \infty$, since $A \in \bb{K}_\infty$ forces $(1+r)^\alpha f_j(r,A) \to 0$. Therefore, $A^{(n)} \to A$ with respect to the Fréchet topology.
\end{proof}

\begin{remark}
	From the norm equivalence Eq.~\eqref{eq: equivalence norms f} and Proposition~\ref{Prop: properties Frechet algebra compact case}, the families of norms $\{|||\cdot|||_{j, \alpha}\}_{\alpha \in \N_0}$, for $j \in \Z$, all induce the same Fréchet topology on $\bb{K}_\infty$. Thus, unless otherwise stated, we will take the localization site to be $0$ and simplify the notation by writing $|||A|||_{\alpha} \; := \; |||A|||_{0, \alpha}.$
\hfill $\blacktriangleleft$
\end{remark}

\subsection{Basis dependence on the Fréchet structure}\label{Subsection: basis dependence}
In constructing the Fréchet $\ast$-algebra $\bb{K}_\infty$, we began by fixing an orthonormal basis $\Psi = \{\psi_j\}_{j \in \Z}$ of the Hilbert space $\s{H}$. It is natural to ask how this choice affects the resulting Fréchet structure. Let $\Phi = \{\varphi_j\}_{j \in \Z}$ be another orthonormal basis. Throughout this subsection, we write $V_{j,r}^{\Psi}$, $\bb{K}_{j,r}^{\Psi}$, $\bb{K}_{j,{\rm loc}}^{\Psi}$, $\bb{K}_\infty^{\Psi}$, $f_0^\Psi$ and $p_0^\Psi$ for the corresponding objects constructed from $\Psi$, and use analogous notation for $\Phi$.

 \smallskip

\noindent Since $\Psi$ and $\Phi$ are orthonormal bases, there exists a unitary operator $W: \s{H} \rightarrow \s{H}$ such that $W(\psi_j) = \varphi_j$ for every $j \in \Z$. In particular,
$$W (V_{j,r}^{\Psi}) \; = \; V_{j,r}^{\Phi} \, , \qquad W \bb{K}_{j,r}^{\Psi}W^* \; = \; \bb{K}_{j,r}^{\Phi},$$
for all $j \in \Z$ and $r \in \N_0$. Consequently, conjugation by $W$ defines a continuous $\ast$-isomorphism
$$T_{\Psi}^{\Phi}: \bb{K}_\infty^{\Psi} \longrightarrow \bb{K}_\infty^{\Phi}, \qquad A \longmapsto WAW^* \; ,$$
with respect to the corresponding Fréchet topologies. However, such an isomorphism does not provide the notion of equivalence that interests us here: an arbitrary unitary change of basis need not preserve which operators are local, or almost-local.

 \smallskip 

\noindent For example, we may choose $\Phi$ so that
$$
\varphi_0 \; = \; c \sum_{n \in \Z}\frac{1}{1+|n|} \, \psi_n,
$$
where $c$ is a normalization constant. In this case, the rank-one projection onto $\varphi_0$ is local with respect to $\Phi$ but not with respect to $\Psi$. Thus,
$$ \bb{K}_{j, {\rm loc}}^{\Phi} \neq \bb{K}_{k,{\rm loc}}^{\Psi} \qquad \text{for all } j, k \in \Z.$$

\noindent In particular, unitary equivalence alone does not capture the locality properties we want to retain.

 \smallskip

With this in mind, a first natural requirement to ask for comparing two such structures is
$$ \bb{K}_{j, {\rm loc}}^{\Phi} \; = \; \bb{K}_{k,{\rm loc}}^{\Psi} \qquad \text{for all } j, k \in \Z \; .$$

\noindent A natural question is whether this equality already implies that the corresponding Fréchet algebras coincide. The answer is negative, as the following example illustrates.

\begin{example}
Consider the strictly increasing function $g: \N \rightarrow \N$ defined by $g(n) = 2^{n!}$. Let $\Phi = \{\varphi_j\}_{j \in \Z}$ be the orthonormal basis obtained from $\Psi$ by setting
$$\varphi_{g(j)} = \psi_{g(j + 1) - 1}, \qquad \varphi_{g(j + 1) - 1} = \psi_{g(j)},$$
for every $j \geq 1$, and leaving the remaining basis vectors unchanged. Since $\Phi$ is obtained by permuting the elements of $\Psi$, the corresponding algebras of local operators coincide. Nevertheless, the associated Fréchet algebras are different. To see this, consider
$$A \; = \; \sum_{n = 1}^\infty \frac{1}{g(n)^n} \ket{\varphi_{g(n)}} \bra{\varphi_{g(n)}} \; .$$
Clearly $A \in \bb{K}$. Now, observe that $A$ belongs to $\bb{K}_\infty^\Phi$. Indeed, for any $\alpha \in \N_0$, one has
\begin{align*}
(1 + r)^\alpha p_0^\Phi(r, A) \; &= \; (1 + r)^\alpha \sup_{g(n) > r} \frac{1}{g(n)^n} \\
& \leqslant \; \sup_{n \geq 1} g(n)^{\alpha - n} \\
& < \; \infty \; .
\end{align*}
Nevertheless, a simple computation shows that $A \notin \bb{K}_\infty^\Psi$. To see this, first observe that $A$ can be expressed as
$$A \; = \; \sum_{n = 1}^\infty \frac{1}{g(n)^n} \ket{\psi_{g(n + 1) - 1}} \bra{\psi_{g(n + 1) - 1}}. $$
Set $r_n := g(n + 1) - 2$. We then have
\begin{align*}
(1 + r_n) p_0^\Psi (r_n, A) \; &= \; \frac{g(n + 1) - 1}{g(n)^n}  \; = \; g(n) - g(n)^{-n} \; , 
\end{align*} 
which tends to $\infty$ as $n \to \infty$. 
\hfill $\blacktriangleleft$
\end{example}

We are therefore led to a stronger notion of equivalence: we require not only that the two constructions have the same local algebra, but also that $\bb{K}_\infty^{\Psi} = \bb{K}_\infty^{\Phi}$ as subalgebras of $\bb{K}$, with the identity map continuous in both directions for the corresponding Fréchet topologies. The following proposition gives a sufficient condition for this form of equivalence. 

\begin{proposition}
Let $\Psi$ and $\Phi$ be two orthonormal bases of $\s{H}$. Suppose that there exist strictly increasing polynomial functions $p, q: \N_0 \rightarrow \N_0$ such that, for every $r\in\N_0$,
$$V_{0, r}^\Psi \; \subseteq \; V_{0, p(r)}^\Phi, \qquad V_{0, r}^\Phi \; \subseteq \; V_{0, q(r)}^\Psi.$$
Then $\bb{K}_{0, {\rm loc}}^\Psi = \bb{K}_{0, {\rm loc}}^\Phi$. Moreover, their associated Fréchet structures are equivalent, in the sense that $\bb{K}_\infty^\Psi = \bb{K}_\infty^\Phi$ and the identity map is continuous in both directions for the corresponding Fréchet topologies.
\end{proposition}

\begin{proof}
The equality $\bb{K}_{0, {\rm loc}}^\Psi = \bb{K}_{0, {\rm loc}}^\Phi$ follows directly from the assumptions. 

\noindent Set $d_p := \operatorname{deg} p$ and, for each $r \in \N_0$, define $I_r := \{p(r), \ldots, p(r + 1)-1\}$. Let $A \in \bb{K}_\infty^\Psi$ and fix $\alpha \in \N_0$. Since $f_0^\Phi(\cdot, A)$ is decreasing and
$\bb{K}_{0,r}^\Psi \subseteq \bb{K}_{0,p(r)}^\Phi$, we obtain
\begin{align*}
\sup_{s \in I_r}(1 + s)^\alpha f_0^\Phi(s, A) \; &\leqslant \; p(r + 1)^\alpha f_0^\Phi(p(r), A)\\
&\leqslant \; p(r + 1)^\alpha f_0^\Psi(r, A)\\
&\leqslant \; C_\alpha (1 + r)^{d_p \alpha} f_0^\Psi(r,A) \; ,
\end{align*}
where $C_\alpha > 0$ is a constant depending only on $\alpha$ and $p$. On the other hand, since $f_0^\Phi(s, A) \leq ||A||$ for all $s$,
$$\sup_{0 \leq s \leq p(0)}(1 + s)^\alpha f_0^\Phi(s, A) \; \leqslant \; (1 + p(0))^\alpha ||A|| \; .$$
Thus, one has
\begin{align*}
|||A|||_\alpha^\Phi \; &= \; ||A|| + \sup_{s \in \N_0}(1 + s)^\alpha f_0^\Phi(s, A)\\
&= \; ||A|| + \max\left(\sup_{0 \leq s \leq p(0)}(1 + s)^\alpha f_0^\Phi(s, A), \, \sup_{r \in \N_0} \left(\sup_{s \in I_r}(1 + s)^\alpha f_0^\Phi(s, A)\right)\right)\\
&\leqslant \; (2 + p(0))^\alpha ||A|| + C_\alpha \sup_{r \in \N_0}(1 + r)^{d_p \alpha} f_0^\Psi(r, A)\\
&\leqslant \; C_\alpha' |||A|||_{d_p \alpha}^\Psi \; .
\end{align*}
This shows that $\bb{K}_\infty^\Psi \subseteq \bb{K}_\infty^\Phi$, with continuous inclusion for the corresponding Fréchet topologies. Repeating the argument with $q$ in place of $p$ yields the reverse inclusion, together with continuity in the opposite direction. Therefore, the two Fréchet algebras coincide, and the identity map is continuous in both directions.
\end{proof}

\subsection{Description in terms of matrix coefficients}\label{Subsection: matrix coefficients}

We now introduce a new family of functions designed to measure locality in terms of the matrix coefficients with respect to the basis $\Psi$. This new family will induce an equivalent system of norms on the space $\bb{K}_\infty$, which will be particularly useful for understanding continuous derivations on this algebra. Given $j \in \Z$ and $\alpha \in \N_0$, define the norms $||\cdot||_{j, \alpha}$, initially on $\bb{K}_{\rm loc}$, as
\begin{equation*}
	||A||_{j, \alpha} \; := \; \sup_{n,m \in \Z} (1 + |j - n| + |j - m|)^\alpha |A_{n,m}|\;,
\end{equation*}
where $A_{n,m} := \langle \psi_n, A \psi_m \rangle$ are the matrix coefficients of $A$ with respect to $\Psi$. To show that this family of norms generates the same topology on $\bb{K}_\infty$ as the rapidly approximating norms introduced previously, we will require the following well-known technical result (see \cite[Chapter II]{Conway}).
\begin{lemma}[Schur test]\label{lemma: Schur test}
		Let $\mathsf{A} \equiv [\mathsf{A}_{n,m}]_{n,m \in \Z}$ be an infinite matrix with complex entries. Suppose that there exist constants $R, C > 0$ such that
		$$\sup_{n \in \Z} \sum_{m \in\Z} |\mathsf{A}_{n,m}| \;\leqslant\; R\;, \qquad \sup_{m \in \Z} \sum_{n \in\Z} |\mathsf{A}_{n,m}| \;\leqslant\; C\;.$$
		Then $\mathsf{A}$ defines a bounded operator $A$ on $\ell^2(\Z)$ with operator-norm bound $||A||^2 \leqslant RC$.
\end{lemma}
%\begin{proof}
%	Let $v \in \ell^2(\Z)$, and write it in the canonical basis as
%	$$v = \sum_{n \in \Z} v_n e_n.$$
%	For each $n \in \Z$, consider the (formal) expression $w_n := \sum_{m \in \Z} A_{n,m} v_m = \langle e_n, A v \rangle.$ We first show that the series defining $w_n$ is absolutely convergent.  
%By the Cauchy--Schwarz inequality,
%	\begin{align*}
%	\left(\sum_{m} |A_{n,m}| \ |v_m|\right)^2 \; &= \; \left(\sum_{m} |A_{n,m}|^{1/2} \left(|A_{n,m}|^{1/2} |v_m|\right)\right)^2 \\
%	&\leqslant \left( \sum_m |A_{n,m}|\right) \left(\sum_m |A_{n,m}| \ |v_m|^2\right)\\
%	& \leqslant R^2 ||v||^2. 
%	\end{align*}
%	We now prove that $w = (w_n)_{n \in \Z}$ belongs to $\ell^2(\Z)$. Using the estimate above and summing over $n$, we obtain	
%	\begin{align*}
%		||w||^2 = \sum_n |w_n|^2 &\leqslant \sum_n \left(\sum_m |A_{n,m}| \ |v_m|\right)^2 \\
%		&\leqslant \sum_n \left( \sum_{m'}  |A_{n,m'}|\right) \left(\sum_{m} |A_{n, m}| \ |v_{m}|^2\right)\\
%		&\leqslant R \sum_m |v_m|^2 \left(\sum_n |A_{n,m}|\right)\\
%		&\leqslant RC ||v||^2.
%	\end{align*}
%	Thus $w \in \ell^2(\Z)$, the operator $A$ is well defined on $\ell^2(\Z)$, and the estimate above shows that $||A||^2 \leqslant RC$. 
%\end{proof}

 \smallskip

Observe that for any $A \in \bb{K}$ one has the following estimate:
\begin{align*}
	\sum_{m \in \Z} |A_{n,m}|
	\;&=\; \sum_{m \in \Z} (1 + |n - j| + |m - j|)^{\alpha} |A_{n,m}| \, (1 + |n - j| + |m - j|)^{-\alpha} \\
	&\leqslant\; \|A\|_{j,\alpha} \sum_{m \in \Z} (1 + |n - j| + |m - j|)^{-\alpha}
	\;\leqslant\; S_\alpha \, \|A\|_{j,\alpha}\;,
\end{align*}
where
$$
	S_\alpha \;:=\; \sup_{n \in \Z} \sum_{m \in \Z} (1 + |n - j| + |m - j|)^{-\alpha}.
$$
In particular, one has $S_\alpha < \infty$ whenever $\alpha \geq 2$. Hence,
$$
	\sup_{n \in \Z} \sum_{m \in \Z} |A_{n,m}| \;\leqslant\; S_\alpha \|A\|_{j,\alpha}\;.
$$
Similarly,
$$
	\sup_{m \in \Z} \sum_{n \in \Z} |A_{n,m}|\; \leqslant \;S_\alpha \|A\|_{j,\alpha}\;.
$$
By the previous lemma, we therefore obtain
\begin{equation}\label{remark: Operator norm vs matrix norm}
	\|A\|\; \leqslant\; S_\alpha \, \|A\|_{j,\alpha}\;.
\end{equation}
Notice that this estimate remains valid even in the case $\|A\|_{j,\alpha} = \infty$. 

 \smallskip

As with the norms $|||\cdot|||_{j, \alpha}$, the norms $||\cdot||_{j, \alpha}$ are equivalent for different choices of localization site. Indeed, for any $j, k \in \Z$, setting $R = |j - k|$, we have
\begin{equation}\label{eq: equivalence matrix norm}
	\frac{1}{(1 + 2R)^\alpha} ||A||_{j, \alpha} \; \leqslant \; ||A||_{k, \alpha} \; \leqslant \; (1 + 2R)^\alpha||A||_{j, \alpha}\;.
\end{equation} 
As before, unless otherwise stated, we localize around $0$ and simplify the notation by writing $||\cdot||_{\alpha} := ||\cdot||_{0, \alpha}$.

\begin{proposition}\label{Proposition: Equivalence norms compact algebra}
	The families of norms $\{|||\cdot|||_{\alpha}\}_{\alpha \in \N_0}$ and $\{||\cdot||_{\beta}\}_{\beta \in \N_0}$ are equivalent. Therefore, the family $\{\|\cdot\|_{\beta}\}_{\beta \in \N_0}$ generates the Fr\'echet topology on $\bb{K}_\infty$.
\end{proposition}

\begin{proof}
Let $A \in \bb{K}_\infty$. For  $n,m \in \Z$ with $(n,m)\neq(0,0)$, define $r_{n,m} := \max\{|n|,|m|\} - 1$. Observe that for any $B \in \bb{K}_{0,r_{n,m}}$ we have $B_{n,m} = 0$. Thus, for each matrix coefficient one obtains
\begin{equation*}
	|A_{n,m}|\; =\; |(A-B)_{n,m}|\; \leqslant\; \|A-B\|\;,
\end{equation*}
which implies $|A_{n,m}| \leqslant f_0(r_{n,m},A)$.
Next, note that
$$
1 + |n| + |m| \;\leqslant\; 3 + 2 r_{n,m} \;\le\; 3(1 + r_{n,m})\;.
$$
Therefore,
$$
(1 + |n| + |m|)^\beta |A_{n,m}| \; \leqslant \; 3^\beta (1 + r_{n,m})^\beta f_0(r_{n,m},A)\;.
$$
Since $|A_{0,0}| \leq ||A||$, one gets:
\begin{equation}\label{eq0: proofEquivNorms}
	\begin{aligned}
\|A\|_{\beta} \; &= \; \sup_{n,m \in \Z}(1 + |n| + |m|)^\beta |A_{n,m}|\\ & \leqslant \; ||A|| +  3^\beta \sup_{r \in \N_0} (1 + r)^\beta f_0(r,A) \; \leqslant \; 3^\beta \ |||A|||_{\beta}\;.
\end{aligned}
\end{equation}

To prove the reverse inequality, we use the estimate \eqref{remark: Operator norm vs matrix norm}. For $\alpha \geqslant 2$,
\begin{equation}\label{eq1: proofEquivNorms}
	p_0(r,A) \; = \; \|A - P_{0,r} A P_{0,r}\| \; \leqslant \; S_\alpha \|A - P_{0,r} A P_{0,r}\|_{0,\alpha}\;.
\end{equation}
Observe that $(A - P_{0,r} A P_{0,r})_{n,m} = 0$ whenever $|n| \leq r$ and $|m| \leq r$. Hence,
\begin{equation}\label{eq2: proofEquivNorms}
	\|A - P_{0,r} A P_{0,r}\|_{0,\alpha} \; = \; \sup_{\substack{n,m \in \Z \\ \max\{|n|,|m|\} > r}} (1+|n|+|m|)^\alpha |A_{n,m}|\;.
\end{equation}
Since the supremum in \eqref{eq2: proofEquivNorms} is taken over indices with $\max\{|n|,|m|\} > r$, it follows that
$$
1+r \;\leqslant\; 1+|n|+|m|\;.
$$
Multiplying \eqref{eq1: proofEquivNorms} by $(1+r)^\beta$ and taking $\alpha = 2$, we obtain
\begin{align*}
(1 + r)^\beta p_0(r,A) \;&\leqslant\; S_2 (1 + r)^\beta \sup_{\max\{|n|,|m|\} > r} (1 + |n| + |m|)^2 |A_{n,m}|\\
&\leqslant\; S_2 \sup_{\max\{|n|,|m|\} > r} (1 + |n| + |m|)^{2 + \beta} |A_{n,m}|\\
&\leqslant S_2 \|A\|_{2 + \beta}\;.
\end{align*}
Taking the supremum over $r \in \mathbb{N}_0$ gives
\begin{equation}\label{eq3: proofEquivNorms}
	|||A|||_{\beta} \; = \; ||A|| + \sup_{r}(1 + r)^\beta p_0(r,A) \; \leqslant \; (1+ S_2) \, \|A\|_{2+\beta}\;.
\end{equation}

\noindent Combining Eq.~\eqref{eq0: proofEquivNorms} and Eq.~\eqref{eq3: proofEquivNorms} yields the equivalence of the two families of norms as claimed.
\end{proof}

\subsection{The Dirac derivation}\label{Subsection: Dirac derivation}

We now introduce an unbounded self-adjoint operator on the Hilbert space $\s{H}$, defined in a natural way with respect to the basis $\Psi$. On the dense subspace
$$
\s{H}^\Psi_0 \;:=\; {\rm span}\{\psi_n\;|\; \ n \in \Z\}\; \subset\; \s{H}
$$
define the \emph{Dirac operator} $D$ by prescribing its action on the basis elements,
$$D(\psi_n) \;:=\; n \psi_n, \qquad n \in \Z.$$
It extends to a (unique) self-adjoint operator $D$ on $\s{H}$ with domain $D(D) \supseteq \s{H}^\Psi_0$. Recall that a linear map $\delta: D(\delta) \subseteq \bb{K} \rightarrow \bb{K}$, defined on a subalgebra $D(\delta)$, is called a \emph{derivation} if it satisfies the Leibniz rule
$$\delta(AB) \; = \; \delta(A)B + A \delta(B)\;, \qquad \forall\, A,B \in D(\delta)\;.$$
The operator $D$ induces a derivation on the algebra of local operators via the commutator:
\begin{equation*}
    \delta_D(A) \; := \; \ii [D, A]\;, \qquad A \in \bb{K}_{\rm loc}\;.
\end{equation*}
It is straightforward to check that $\delta_D(A)$ is again a local operator, so that $\delta_D: \bb{K}_{\rm loc} \rightarrow \bb{K}_{\rm loc}$ is a well-defined derivation. In the following result we show that $\delta_D$ extends to a continuous derivation on $\bb{K}_\infty$ and we will refer to it as the \emph{Dirac derivation}. Recall that a linear map $T: \bb{K}_\infty \rightarrow \bb{K}_\infty$ is said to be continuous if for every $\alpha \in \N_0$, there exist $C_\alpha > 0$ and $\beta \in \N_0$ such that
$$|||T(A)|||_{\alpha} \;\leqslant\; C_\alpha |||A|||_{\beta}\;,$$
for any $A \in \bb{K}_\infty$.

\begin{proposition}\label{prop: continuity delta D}
    The derivation $\delta_D$ extends to a continuous derivation on the Fréchet $\ast$--algebra $\bb{K}_\infty$.
\end{proposition}
\begin{proof}
    Given $A \in \bb{K}_{\rm loc}$, a direct computation shows that
    $$(\delta_D(A))_{n,m}\; =\; \ii \,(n - m)\, A_{n,m}\;.$$
    Therefore, for any $j \in \Z$,
    \begin{align*}
        ||\delta_D(A)||_{j, \alpha} \; &= \; \sup_{n,m \in \Z} (1 + |n - j| + |m - j|)^\alpha |(\delta_D(A))_{n,m}|\\
        &=\; \sup_{n,m \in \Z} (1 + |n - j| + |m - j|)^\alpha |n - m| \ |A_{n,m}|\\
        &\leqslant\; \sup_{n,m \in \Z} (1 + |n - j| + |m - j|)^{\alpha + 1} |A_{n,m}|\\
        &= \;||A||_{j, \alpha + 1}\;.
    \end{align*}
    Since $\bb{K}_{\mathrm{loc}}$ is dense in $\bb{K}_\infty$ in the Fréchet topology, this estimate implies that $\delta_D$ extends uniquely to a continuous linear map on $\bb{K}_\infty$. Finally, the derivation identity $\delta_D(AB) = \delta_D(A) B + A \delta_D(B)$ extends from $\bb{K}_{\mathrm{loc}}$ to $\bb{K}_\infty$ by density and continuity of the product. Thus, the extension is a continuous derivation.
    \end{proof}

 \smallskip

We conclude this subsection by constructing a one-parameter group of continuous automorphisms on the algebra $\bb{K}_\infty$ arising from the operator $D$. Since $D$ is self-adjoint, it generates a strongly continuous one-parameter unitary group on the Hilbert space $\s{H}$,
$$U_D(t) \; := \; e^{\ii tD}, \qquad t \in \R\;.$$
One may then consider the strongly continuous one-parameter group of automorphisms $\rho_D: \R \to {\rm Aut}(\bb{K})$ defined by 
\[\rho_D(t)(A)\; :=\; {\rm Ad}_{U_D(t)}(A)\; =\; U_D(t) A U_D(t)^*\;.
\] 
The next result will show that $\rho_D(t)(\bb{K}_\infty) \subseteq \bb{K}_\infty$. Moreover,  $\rho_D$ restricts to a strongly continuous one-parameter group with respect to the Fréchet topology of $\bb{K}_\infty$.

\begin{proposition}\label{Prop: continuity rhoD}
    For each $t \in \R$, the map $\rho_D(t)$ defines a continuous automorphism of the Fréchet algebra $\bb{K}_\infty$. Moreover, $\rho_D: \R \to {\rm Aut}(\bb{K}_\infty)$ is a strongly continuous one-parameter group of automorphisms. Finally, for every $A \in \bb{K}_\infty$, the curve $t \mapsto \rho_D(t)(A)$ is the unique solution to the differential equation
    \begin{equation}\label{eq0: Prop continuity rhoD}
        \frac{d}{dt}\rho_D(t)(A)\; =\; \rho_D(t)(\delta_D(A))\;, \qquad \rho_D(0)(A) \;=\; A\;.
    \end{equation}
\end{proposition}
\begin{proof}
    We begin by showing that for any $t \in \R$ and any $A \in \bb{K}_\infty$, one has $\rho_D(t)(A) \in \bb{K}_\infty$. Since $U_D(t)$ is diagonal in the basis $\Psi$, with diagonal entries $e^{\ii tn}$, one computes
    \begin{equation*}
    	||\rho_D(t)(A)||_{ \alpha} \;=\; \sup_{n,m \in \Z} (1 + |n| + |m|)^\alpha |e^{\ii t(n - m)}A_{n,m}| \;=\; ||A||_{\alpha}\;.
    \end{equation*}
    Thus, for each $t \in \R$, we have that $\rho_D(t)(\bb{K}_\infty) \subseteq \bb{K}_\infty$. Moreover, the above estimation shows that $\rho_D(t)$ is a continuous $*$-homomorphism with respect to the Fréchet topology. Since $\rho_D(-t)$ is its inverse, it follows that $\rho_D(t)$ is a continuous automorphism.
To show that $\rho_D$ defines a strongly continuous one-parameter group, we must prove that for every $A \in \bb{K}_\infty$, the map $t \mapsto \rho_D(t)(A)$ is continuous. Let $s, t \in \R$. Using the estimate $|e^{\ii r} - 1| \leqslant |r|$, we obtain
    \begin{align*}
        ||\rho_D(t)(A) - \rho_D(s)(A)||_{\alpha} \;&=\; \sup_{n,m \in \Z} (1 + |n| + |m|)^\alpha |e^{\ii (t - s)(n - m)} - 1| \ |A_{n,m}|\\
        &\leqslant\; |t - s| \sup_{n,m \in \Z} (1 + |n| + |m|)^\alpha |n - m| \ |A_{n,m}|\\
        &\leqslant\; |t - s| \ ||A||_{\alpha + 1}\;.
    \end{align*}
    Since $||A||_{\alpha + 1} < \infty$, strong continuity follows.
Finally, we show that for each $A \in \bb{K}_\infty$, the curve $t \mapsto \rho_D(t)(A)$ is the unique solution of the differential equation \eqref{eq0: Prop continuity rhoD}. Using the Taylor remainder estimate 
    $$|e^{\ii x} - 1 - \ii x| \;\leqslant\; \frac{x^2}{2}\;, \qquad x \in \R\;,$$
    we obtain
    \begin{align*}
        &\left|\left|\frac{\rho_D(t + h)(A) - \rho_D(t)(A)}{h} - \rho_D(t)(\delta_D(A))\right|\right|_{ \alpha} \\
        &\quad\qquad=\; \sup_{n,m \in \Z} (1 + |n| + |m|)^\alpha \left|\frac{e^{\ii h(n - m)} - 1}{h} - \ii (n - m)\right| |A_{n,m}|\\
        &\quad\qquad\leqslant\; \frac{|h|}{2} \sup_{n,m \in \Z} (1 + |n| + |m|)^\alpha |n - m|^2 |A_{n,m}|\\
        &\quad\qquad\leqslant\; \frac{|h|}{2} ||A||_{\alpha + 2}\;.
    \end{align*}
    Since $||A||_{ \alpha + 2} < \infty$, the right-hand side tends to $0$ as $h \rightarrow 0$, proving that
    $$\frac{d}{dt}\rho_D(t)(A) \;= \;\rho_D(t)(\delta_D(A))$$
   in the Fréchet topology. Since $\rho_D(0)(A) = A$, the curve satisfies the initial condition. To prove uniqueness, let $B: \mathbb{R} \to \bb{K}_\infty$ be any continuously differentiable solution with $B(0) = A$. For fixed $t \in \mathbb{R}$, consider $F(s) := \rho_D(-s)(B(s))$. Differentiating and using the group property gives 
   $$F'(s) \; = \; -\rho_D(-s)(\delta_D(B(s))) + \rho_D(-s)(B'(s)) \; = \; 0 \; .$$
Thus, $F(s)$ is constant and equals $F(0) = A$. Consequently, $B(s) = \rho_D(s)(A)$, which proves uniqueness.
\end{proof}

\section{Fr\'echet--continuous derivations}\label{Section: Frechet continuous derivations}

Having established the continuity of the derivation $\delta_D$ and its associated one-parameter group, we now aim to characterize all continuous derivations $\delta$ on the Fréchet algebra $\bb{K}_\infty$. This characterization has been described in~\cite{Klimek}, where the authors worked with an orthonormal basis indexed by $\N_0$. In our setting, however, the orthonormal basis $\Psi$ is indexed by $\Z$. Although the statements of their results carry over with only minor modifications, this change requires some adjustments to the proofs. We therefore include these proofs to keep the exposition self-contained and to account for the differences between the two settings.

 \smallskip

\noindent An important consequence of this characterization is an analogue of the classical result concerning bounded derivations on $\bb{K}$. More precisely, we shall see that every Fréchet-continuous derivation $\delta:\bb{K}_\infty \rightarrow \bb{K}_\infty$ admits an implementation by an operator that is continuous on a suitable subspace of $\s{H}$ endowed with a Fréchet topology (see Definition~\ref{Def: smooth HS} and Proposition~\ref{Proposition: Domain Bdelta}).

\subsection{Covariant derivations} We begin by recalling two definitions from~\cite{Klimek}.

\begin{definition}\label{Def: n-covariant}
	A continuous derivation $\delta: \bb{K}_\infty \rightarrow \bb{K}_\infty$ is said to be $n$-covariant, for $n \in \Z$, if for all $t \in \R$ and all $A \in \bb{K}_\infty$ one has
	$$\rho_D(-t) \circ \delta \circ \rho_D(t) (A) \; = \; e^{-\ii n t} \delta(A).$$
	If $n = 0$, we say that $\delta$ is invariant. 
\end{definition}
\begin{definition}\label{Def: Fourier components}
	Let $\delta: \bb{K}_\infty \rightarrow \bb{K}_\infty$ be a continuous derivation. The \textit{n-th Fourier component} of $\delta$ is defined by
	$$\delta_n(A) \; := \; \frac{1}{2\pi} \int_{0}^{2\pi} e^{\ii n t} \rho_D(-t) \circ \delta \circ \rho_D(t) (A) \, dt.$$
\end{definition}
The following Proposition is a restatement of \cite[Proposition 4.1 and Lemma 4.2]{Klimek}
\begin{proposition}\label{Lemma: decaying Fourier coefficients}
	Let $\delta: \bb{K}_\infty \rightarrow \bb{K}_\infty$ be a continuous derivation. Then, for each $n \in \Z$, its Fourier component $\delta_n$ defines a continuous $n$-covariant derivation on $\bb{K}_\infty$. 
	
	\noindent Moreover, for every $\alpha,\gamma \in \mathbb{N}_0$, there exist $\beta \equiv \beta_{\alpha, \gamma} \in \N_0$ and $C \equiv C_{\alpha, \gamma} > 0$ such that 	
	$$ |n|^\gamma \, ||\delta_n(A)||_{\alpha} \; \leqslant \; C \, ||A||_{\beta} $$
	for all $A \in \bb{K}_\infty$ and $n \in \Z$.
\end{proposition}
%\begin{proof}
	%Since $\delta$ is continuous, and since $\rho_D(t)$ preserves all norms of the Fr\'echet topology (see Proposition~\ref{Prop: continuity rhoD}), it follows that the map
	%$$F_t(A) \;:=\; e^{\ii nt}\,\rho_D(-t)\bigl(\delta(\rho_D(t)(A))\bigr)	$$
%is continuous in $t \in [0,2\pi]$ and takes values in the complete locally convex space $\bb{K}_\infty$. By standard results on vector-valued integration (see \cite[\S 3]{RudinFunctional}, in particular Exercise 23), the integral
%$$\delta_n(A) \; := \; \frac{1}{2\pi}\int_0^{2\pi} F_t(A)\, dt$$
%is well defined and yields a continuous linear map $\bb{K}_\infty \rightarrow \bb{K}_\infty$. Using that $\rho_D(t)$ is a $*$--automorphism and that $\delta$ satisfies Leibniz rule, a straightforward computation shows that $F_t(AB) = F_t(A)B + AF_t(B)$, and therefore $\delta_n$ is a derivation. It remains to show that $\delta_n$ is $n$-covariant. For any $s\in\R$,
%\begin{align*}
%\rho_D(-s) \, \delta_n (\rho_D(s)(A)) \; &= \; \frac{1}{2\pi}\int_0^{2\pi} e^{\ii nt}\,\rho_D(-(t + s))\, \delta(\rho_D(t + s)(A))\, dt \\
%&=\; \frac{1}{2\pi} \int_s^{2\pi + s} e^{\ii n(u-s)} \, \rho_D(-u)\,\delta(\rho_D(u)(A))\, du \\
%&=\; e^{-\ii ns} \, \delta_n(A)\;,
%\end{align*}
%where we used the change of variables $u := t + s$ and the periodicity of the integrand to shift the interval of integration. This is exactly the condition that $\delta_n$ is $n$-covariant
%\end{proof}
\begin{proof}
	The proof of the first statement is the same as in~\cite[Proposition 4.1]{Klimek}, so we omit it and focus on the second assertion.
	
	 \smallskip
	
	By continuity of $\delta$, it follows that for any $\alpha \in \N_0$ there exists $C_\alpha > 0$ and $\beta \in \N_0$ such that 
	$$||\delta(A)||_{\alpha}\; \leqslant\; C_\alpha \, ||A||_{\beta} \; .$$
	Observe that the same estimate holds for each Fourier component $\delta_n$, uniformly in $n \in \Z$. Indeed, the integral defining $\delta_n$ is well defined with respect to the Fréchet topology, and using the continuity estimate for $\delta$, we obtain
	\begin{align*}
		||\delta_n(A)||_{\alpha} \; &\leqslant\; \frac{1}{2\pi} \int_0^{2\pi} ||\rho_D(-t)\bigl(\delta(\rho_D(t)(A))\bigr)||_{\alpha} \, dt \\
		&\leqslant \;  \frac{C_\alpha}{2\pi} \int_{0}^{2\pi} ||\rho_D(t)(A)||_{\beta}\, dt = C_\alpha \, ||A||_{\beta}\;,
	\end{align*}
	for all $A \in \bb{K}_\infty$. 
	
	\noindent Now, fix $A$ in the algebra. By Proposition~\ref{Prop: continuity rhoD}, the map $t \mapsto \rho_D(-t)\bigl(\delta(\rho_D(t)(A))$ is continuously differentiable in the Fr\'echet sense. Differentiating, we obtain
	$$\frac{d}{dt} \left(\rho_D(-t)\circ \delta \circ \rho_D(t)(A)\right) \; = \; \rho_D(-t) \circ [\delta, \delta_D] \circ \rho_D(t) (A)\;,$$
	where $[\delta, \delta_D](A) = \delta(\delta_D(A)) - \delta_D(\delta(A))$. Therefore,
	\begin{align*}
		\ii n \delta_n(A) \; &= \; \frac{1}{2\pi} \int_{0}^{2\pi} in e^{\ii n t} \rho_D(-t)\circ \delta \circ \rho_D(t)(A) \, dt \\
		&= \; \frac{1}{2\pi} \int_{0}^{2\pi} \frac{d}{dt} (e^{\ii n t}) \rho_D(-t)\circ \delta \circ \rho_D(t)(A) \, dt \\
		&= \; - \frac{1}{2\pi} \int_{0}^{2\pi}  e^{\ii n t} \frac{d}{dt}\left(\rho_D(-t)\circ \delta \circ \rho_D(t)(A)\right) \, dt\\
		&= \; - ([\delta, \delta_D])_n (A)\; ,
	\end{align*}
	where, in the third equality, we have used integration by parts and the $2\pi$-periodicity of $\rho_D$. Since both $\delta$ and $\delta_D$ are continuous, their commutator is continuous as well. Hence the same estimate that bounds $\|[\delta,\delta_D](A)\|_{\alpha}$ also bounds $\|[\delta,\delta_D]_n(A)\|_{\alpha}$. Thus, given $\alpha \in \N_0$, there exists $C > 0$ and $\beta \in \N_0$ such that
	$$|n| \, ||\delta_n(A)||_{\alpha}\; \leqslant\; C \, ||A||_{\beta}\;. $$
	An inductive argument now yields the general case for any power of $n$. Indeed,	
	$$
	\begin{aligned}
	(\ii n)^\gamma \delta_n(A) \;&=\; -(\ii n)^{\gamma- 1} ([\delta, \delta_D])_n (A) \;=\; \ldots\\
	 &=\; (-1)^\gamma ([[\ldots[ \delta, \delta_D], \delta_D] \ldots ])_n(A)\;,
	 \end{aligned}
	 $$
	where the iterated commutator is taken $\gamma$-times.
	\end{proof}

The Fourier components introduced above allow us to use several familiar results from classical Fourier analysis. In particular, some results for periodic functions $f : \R \to \C$ admitting a Fourier series decomposition extend to continuous periodic mappings $F : \R \to X$, where $X$ is a complete Hausdorff locally convex topological vector space (in particular, a Fréchet space). In \cite{Fourier}, the authors establish several such results in the context of continuous representations of the circle on $X$. 

 \smallskip

Although in our case the map $F_t$ is not strictly a representation, in \cite[Theorem~2.1]{Fourier} the authors prove a version of Fej\'er's theorem whose proof applies verbatim in our setting. Roughly speaking, this theorem asserts that the Ces\`aro means of the partial sums of the Fourier series,
$$S_j(\delta(A)) \; := \; \sum_{n = -j}^j \delta_n(A),$$
converges, in the Fr\'echet topology, to $\delta(A)$. Moreover, by Proposition~\ref{Lemma: decaying Fourier coefficients}, given $A \in \bb{K}_\infty$ and $\alpha \in \N_0$, the series
$$
\sum_{n \in \Z} \|\delta_n(A)\|_{\alpha}
$$
is convergent. 
Hence, the Fourier series $\sum_{n \in \Z} \delta_n(A)$ converges absolutely, in the Fr\'echet sense, to $\delta(A)$.

 \smallskip

With this in mind, we now turn to the description of continuous $n$-covariant derivations. Before doing so, we introduce some notation. Let $Q_{n,m} := \ketbra{\psi_n}{\psi_m}$ denote the matrix units associated with the basis $\Psi$. Every operator $A \in \bb{K}_\infty$ can then be expressed as
\begin{equation}\label{eq: formal series element K infty}
	A \;=\; \sum_{n,m \in \Z} A_{n,m}\, Q_{n,m},
\end{equation}
where the series converges absolutely with respect to the norms $\|\cdot\|_\alpha$ for every $\alpha \in \N_0$. We also use the notation $\chi_{r,s}$ for the \emph{Kronecker delta}, rather than the more common $\delta_{r,s}$, to avoid confusion with the notation for derivations.

 \smallskip

\noindent The following Lemma shows that a formal series as in Eq.~\eqref{eq: formal series element K infty} with rapidly decaying coefficients defines an element of $\bb{K}_\infty$.
\begin{lemma}\label{Lemma: matrix coefficients}
	Let $\mathsf{A} = \{\mathsf{A}_{n, m}\}_{n, m \in \Z}$ be a family of complex numbers such that, for every $\alpha \in \N_0$,
	$$ C_\alpha \; := \; \sup_{n, m \in \Z} (1 + |n| + |m|)^\alpha |\mathsf{A}_{n,m}| \; < \; \infty $$
	Then the series
	$$A \; := \; \sum_{n, m \in \Z} \mathsf{A}_{n, m} Q_{n, m}$$
	converges absolutely with respect to the family of norms $\{\|\cdot\|_\alpha\}_{\alpha \in \N_0}$ and defines an element of $\bb{K}_\infty$. In particular, $||A||_\alpha = C_\alpha$.
\end{lemma}
\begin{proof}
	Fix $\alpha \in \N_0$. By the assumed decay of the coefficients,
	\begin{align*}
		\sum_{n, m \in \Z} ||\mathsf{A}_{n, m} Q_{n, m}||_\alpha \; &= \; \sum_{n, m \in \Z} |\mathsf{A}_{n, m}| (1 + |n| + |m|)^\alpha \\
		&\leqslant \; C_{\alpha + 4}\sum_{n, m \in \Z} \frac{1}{(1 + |n| + |m|)^4} \\
		&\leqslant \; C_{\alpha + 4} \left(\sum_{n \in \Z} \frac{1}{(1 + |n|)^2}\right)^2  \; < \; \infty \; .
	\end{align*}
	Here, the last inequality follows from $(1 + |n| + |m|)^2 \geq (1 + |n|)(1 + |m|)$. Since $\alpha$ was arbitrary, the series converges absolutely with respect to every defining norm $\|\cdot\|_\alpha$. Therefore, its limit $A$ defines a compact operator and, furthermore, it belongs to $\bb{K}_\infty$.
\end{proof}

In the following proposition, which is a restatement of~\cite[Proposition 4.4]{Klimek}, we characterize continuous $n$-covariant derivations in terms of formal commutators. We will return later to the questions of well-definedness and domains for the formal operators appearing in this representation, and specify the conditions under which they can be rigorously defined.

\begin{proposition}\label{Prop: commutator n-covariant}
	Let $\delta: \bb{K}_\infty \rightarrow \bb{K}_\infty$ be an $n$--covariant continuous derivation. Then, there exists a sequence $\{\beta_j\}_{j\in\Z}$ such that, for every $A \in \bb{K}_\infty$, the element $\delta(A)$ is given by the formal commutator
	\begin{equation}\label{eq0: Prop commutator n-covariant}
		\delta(A)\; =\; \left[ \, \sum_{l \in \Z} \beta_l Q_{l, l -n} \, , A \, \right]\;.
	\end{equation}
	Moreover, the sequence $\{\beta_j\}_{j\in\Z}$ is polynomially bounded in the sense that there exist constants $C>0$ and $N\in\mathbb{N}$ such that $|\beta_j| \leq C(1 + |j|)^N$.
	Conversely, any polynomially bounded sequence $\{\beta_j\}_{j \in \Z}$ defines a $n$--covariant continuous derivation through Eq.~\eqref{eq0: Prop commutator n-covariant}.
\end{proposition}

\begin{proof}
We begin with the converse direction. Fix $n \in \Z$, and let $\{\beta_l\}_{l \in \Z}$ be a polynomially bounded sequence. For any $A \in \bb{K}_\infty$, the formal commutator in Eq.~\eqref{eq0: Prop commutator n-covariant} can be expressed as
\begin{equation}\label{eq: Prop commutator n-covariant expansion commutator}
	\sum_{l, s}\beta_l A_{l - n, s} Q_{l, s} - \sum_{l, r}\beta_l A_{r, l} Q_{r, l - n} \; .
\end{equation}
To show that both series define elements of $\bb{K}_\infty$, we apply Lemma~\ref{Lemma: matrix coefficients} to the families $\{\beta_l A_{l - n, s}\}_{l, s \in \Z}$ and $\{\beta_l A_{r,l}\}_{r, l \in \Z}$. Fix $\alpha\in\N_0$. Using the polynomial growth of $\{\beta_l\}_{l \in \Z}$, we obtain the following estimate for the first family:
\begin{equation}\label{eq: Prop commutator n-covariant converse1}
	\begin{aligned}
	(1 + |l| + |s|)^\alpha |\beta_l| \  |A_{l-n,s}| \; & \leqslant \; C(1 + |l|)^N (1 + |l| + |s|)^\alpha |A_{l-n,s}| \\
	 & \leqslant \; C \ (1 + |l| + |s|)^{\alpha + N} |A_{l-n,s}| \\
	 & \leqslant \; 2^{\alpha + N} C \ (1 + |n|)^{\alpha + N}  (1 + |l - n| + |s|)^{\alpha + N} |A_{l-n,s}| \\
	 & \leqslant \; 2^{\alpha + N} C \ (1 + |n|)^{\alpha + N} ||A||_{\alpha + N} \; .
	\end{aligned}
\end{equation}
For the second family, a similar estimate gives:
\begin{equation}\label{eq: Prop commutator n-covariant converse2}
	\begin{aligned}
	(1 + |r| + |l - n|)^\alpha |\beta_l| \  |A_{r,l}| \; & \leqslant \; C(1 + |l|)^N (1 + |r| + |l|)^\alpha (1 + |n|)^\alpha |A_{r,l}| \\
	 & \leqslant \; C \ (1 + |n|)^\alpha (1 + |l| + |r|)^{\alpha + N} |A_{r,l}| \\
	 & \leqslant \; C \ (1 + |n|)^\alpha ||A||_{\alpha + N} \; .
	\end{aligned}
\end{equation}
Hence, both series in Eq.~\eqref{eq: Prop commutator n-covariant expansion commutator} define elements of $\bb{K}_\infty$. Thus, the formal commutator gives a well-defined linear map $\delta_\beta: \bb{K}_\infty \rightarrow \bb{K}_\infty$. Moreover, estimates~\eqref{eq: Prop commutator n-covariant converse1} and~\eqref{eq: Prop commutator n-covariant converse2} yield
$$\left\|\delta_\beta(A)\right\|_\alpha \; \leqslant \; C \, 2^{1 + \alpha + N} (1 + |n|)^{\alpha + N} \, \|A\|_{\alpha + N} \; ,$$
which establishes the Fréchet continuity of $\delta_\beta$. It remains to check the Leibniz rule and $n$-covariance. For the former, let $A \in \bb{K}_{0, R}$ for some $R \in \N_0$. The formal commutator then reduces to
$$\delta_\beta (A) \; = \; \left[\sum_{l = -R - |n|}^{R + |n|}\beta_l Q_{l, l - n}, A\right].$$
Since this is an ordinary commutator of finite-rank operators, $\delta_\beta$ satisfies the Leibniz rule on $\bb{K}_{\rm loc}$. By density of $\bb{K}_{\rm loc}$ in $\bb{K}_\infty$ and continuity of $\delta_\beta$ and multiplication, this identity extends to the entire Fréchet $\ast$-algebra.

\noindent The $n$-covariance property can similarly be verified by a direct computation on $\bb{K}_{\rm loc}$. As before, continuity and density allow us to extend this identity to all of $\bb{K}_\infty$. 

 \smallskip

We now prove the other direction. Let $\delta: \bb{K}_\infty \rightarrow \bb{K}_\infty$ be a Fréchet continuous $n$-covariant derivation. We begin with the case $n = 0$. Notice that
	$$\rho_D(t)(Q_{r,s}) \;=\; e^{it(r - s)}Q_{r,s}\,.$$
	From this identity, one concludes that an element $A \in \bb{K}_\infty$ satisfies $\rho_D(t)(A) = A$ for all $t \in \R$ if and only if $A$ can be written in the form
	$$
		A \;= \;\sum_{r \in \Z} A_{r, r}\, Q_{r, r}\;,
	$$
	i.e., if $A$ is diagonal with respect to the basis $\Psi$. We denote this commutative subalgebra by $\bb{K}_\infty^{\rm diag}$. Since $\delta$ is $0$-covariant, it follows that for any $A \in \bb{K}_\infty^{\rm diag}$,
	$$
		\rho_D(t)(\delta(A)) \; = \; \delta(\rho_D(t)(A)) \; = \; \delta(A)\;,
	$$
	and therefore $\delta(A) \in \bb{K}_\infty^{\rm diag}$. In particular, for the orthogonal projection $Q_{k,k} \in \bb{K}_\infty^{\rm diag}$ one has
	$$
		\delta(Q_{k,k}) \; = \; \delta(Q_{k,k}^2) \; = \; \delta(Q_{k,k}) Q_{k,k} + Q_{k,k}\delta(Q_{k,k}) \; = \; 2\, Q_{k,k} \, \delta(Q_{k,k})\;.
	$$
	This is equivalent to $(\mathbf{1}-2 Q_{k,k}) \, \delta(Q_{k,k}) = 0,$ which forces $\delta(Q_{k,k}) = 0$. Hence, $\delta\big|_{\bb{K}_\infty^{\rm diag}} = 0$.
	
	\noindent We next consider the action of $\delta$ on the off-diagonal matrix units $Q_{k, l}$ for $k \neq l$. Since $Q_{k,k} Q_{k,l} = Q_{k,l}$ and $Q_{k,l} Q_{l,l} = Q_{k,l}$, one gets
	$$Q_{k,k} \delta(Q_{k,l}) \;=\; \delta(Q_{k,l})\;, \qquad \delta(Q_{k,l}) Q_{l,l}\; =\; \delta(Q_{k,l}) \; .$$
	These identities imply that $\delta(Q_{k,l})$ must be a scalar multiple of $Q_{k,l}$. Hence, $\delta(Q_{k,l}) = b_{k,l} \, Q_{k,l}$ for a family of scalars $\{b_{k,l}\}_{k,l\in\Z}$.
	
	\noindent To determine the corresponding sequence of $\beta_j$, consider the relation
	\begin{equation}\label{eq1: Prop commutator n-covariant}
		Q_{k,l} Q_{i,j} \; = \; \chi_{l,i}\, Q_{k,j}\;.
	\end{equation}
	Applying $\delta$ yields
	$$(b_{k,l} + b_{i,j})\chi_{l,i}\, Q_{k,j} \; = \; b_{k,j}\ \chi_{l,i}\, Q_{k,j}\;.$$
	Setting $i = l$ we obtain the relation $b_{k,l} = b_{k,j} - b_{l,j}$ for all $j \in \Z$. Define $\beta_k = b_{k,0}$. Then, the sequence $\{\beta_k\}_{k \in \Z}$ determines all the coefficients $b_{k,l}$, and hence the derivation $\delta$. In particular, 
	$\delta(Q_{k,l}) = (\beta_k - \beta_l) Q_{k,l}$. 
	Therefore, over $\bb{K}_{\rm loc}$, $\delta$ can be expressed as the formal commutator
	\begin{equation}\label{eq0: Prop commutator n-covariant Formal commutator}
		\delta(A) \; = \; \left[ \, \sum_{k \in \Z} \beta_k Q_{k,k}, A\,\right] \; .
	\end{equation}  
	
	\noindent Now, let us consider the case $n \neq 0$. As above, we go through the matrix units $Q_{k,l}$. Using that $\delta$ is $n$-covariant, one has
	\begin{equation}\label{eq2: Prop commutator n-covariant}
		\rho_D(t)(\delta(Q_{k,l})) \; = \; e^{int} \delta(\rho_D(t)(Q_{k,l})) \; = \; e^{i(n + k - l)t} \delta(Q_{k,l})\;.
	\end{equation}
	Since $\delta(Q_{k,l})$ belongs to $\bb{K}_\infty$, we can expressed it using the matrix units as
	$$\delta(Q_{k,l}) \; = \; \sum_{r,s \in \Z} b_{r,s}(k,l) Q_{r,s}\;,$$
	which implies
	\begin{equation}\label{eq3: Prop commutator n-covariant}
		\rho_D(t)(\delta(Q_{k,l})) \; = \; \sum_{r,s \in \Z} e^{i(r - s)t} b_{r,s}(k,l) Q_{r,s}\;.
	\end{equation}
Comparing Eq.~\eqref{eq2: Prop commutator n-covariant} and Eq.~\eqref{eq3: Prop commutator n-covariant}, we conclude that $b_{r,s}(k,l) = 0$ for those $r,s$ such that $r - s \neq k - l + n$. In other words,
\begin{equation}\label{eq4: Prop commutator n-covariant}
	\delta(Q_{k,l}) \; = \; \sum_{\substack{r,s \in \Z \\ r - s = k - l + n}} b_{r,s}(k,l) Q_{r,s}\;.
\end{equation}
	Applying $\delta$ to Eq.~\eqref{eq1: Prop commutator n-covariant} and using the expression Eq.~\eqref{eq4: Prop commutator n-covariant}, one gets
	\begin{equation}\label{eq5: Prop commutator n-covariant}
		\sum_{\substack{r,s \in \Z \\ r - s = k - l + n}} b_{r,s} (k,l) Q_{r,s} Q_{i,j} \;+\; \sum_{\substack{r',s' \in \Z \\ r' - s' = i - j + n}}  b_{r',s'} (i,j) Q_{k,l} Q_{r',s'} = \chi_{l,i} \delta(Q_{k,j})\;.
	\end{equation}	
	From this equation we distinguish two cases.

\noindent\textbf{Case (i) - \(l\neq i\).}  
Set
$$
r_1 \;:=\; i+k-l+n\;,\qquad r_2 \;:=\; l-i+j-n\;.
$$
Hence Eq.~\eqref{eq5: Prop commutator n-covariant} reduces to
$$
b_{r_1,i}(k,l)\,Q_{r_1,j} \;+\; b_{l,r_2}(i,j)\,Q_{k,r_2} \;=\; 0\;.
$$
If we further assume the special relation $i - l + n = 0$, then $r_1 = k, r_2 = j,$ and the previous identity becomes
$$
\bigl(b_{k,l - n}(k,l) + b_{l,j}(l - n,j) \bigr) \, Q_{k,j} \; = \; 0\;.
$$
Therefore $b_{l,j}(l - n,j) =  -b_{k,l - n}(k,l).$ Observe that the left-hand side does not depend on $k$, while the right-hand side does not depend on $j$. We therefore introduce the notation
$
\beta_l := b_{l,0}(l - n,0)
$, 
and, by the independence just observed, conclude that
$
b_{l,j}(l - n,j)= \beta_l = -b_{k,l - n}(k,l)
$
for all $k,j \in \Z$.

\noindent\textbf{Case (ii) - $i = l$.}  
Putting $i = l$ in Eq.~\eqref{eq5: Prop commutator n-covariant} and using the identification above yields, after a short computation, the explicit formula for the action of $\delta$ on matrix units:
\begin{equation}\label{eq: Prop commutator n-covariant action over units}
		\delta(Q_{k,j}) \; = \; \beta_{k + n}\, Q_{k + n,j} \; - \; \beta_j \, Q_{k,j - n} \; .
\end{equation}
Hence, for all $A \in \bb{K}_{\rm loc}$, one gets
\begin{equation}\label{eq1: Prop commutator n-covariant Formal commutator}
	\delta (A) \;= \;\left[ \, \sum_{l \in \Z} \beta_l Q_{l, l -n} \, , A \, \right] \; .
\end{equation}

\noindent We have thus obtained the formal commutator representations~\eqref{eq0: Prop commutator n-covariant Formal commutator} and~\eqref{eq1: Prop commutator n-covariant Formal commutator} on $\bb{K}_{\rm loc}$, for $n = 0$ and $n \neq 0$, respectively. It remains to establish that the coefficients ${\beta_l}_{l \in \Z}$ are polynomially bounded and that these representations extend to all of $\bb{K}_\infty$. The latter, however, follows from the former by the converse direction proved above. Indeed, once the polynomial bound is established, the corresponding formal commutators define Fréchet-continuous maps on $\bb{K}_\infty$, which coincide with $\delta$ by density of $\bb{K}_{\rm loc}$. We therefore turn to the polynomial growth of the coefficients.

\noindent To establish that the coefficients $\{\beta_l\}_{l \in \Z}$ are polynomially bounded, we use the Fréchet continuity of $\delta$. Evaluating Eq.\eqref{eq: Prop commutator n-covariant action over units} at $Q_{0,l}$, we obtain
$$\delta(Q_{0,l}) \; = \; \beta_n Q_{n, l} - \beta_l Q_{0, l - n} \; ,$$
which also holds for $n = 0$. Recall that 
$$||A|| \leq |||A|||_0 \leq (1 + S_2)||A||_2,$$ where the second inequality follows from estimate~\eqref{eq3: proofEquivNorms}. Applying the reverse triangle inequality and using the Fréchet continuity of $\delta$, one has
\begin{align*}
\big| |\beta_n| - |\beta_l| \big| \; &= \; \left| \|\beta_n Q_{n, l}\| - \|\beta_l Q_{0, l - n}\|\right| \\
&\leqslant \; (1 + S_2) \|\delta(Q_{0, l})\|_2 \\
&\leqslant \; D_2 (1 + S_2) ||Q_{0, l}||_N \\
& = \;   D_2 (1 + S_2) (1 + |l|)^N \; ,
\end{align*}
where $D_2 > 0$ and $N \in \N_0$ arise from the continuity of $\delta$. Consequently,
$$|\beta_l| \; \leqslant \; |\beta_n| + D_2 (1 + S_2) (1 + |l|)^N \; ,$$
which establishes the required polynomial bound. Together with the extension argument given above, this completes the proof.
\end{proof}

\subsection{Generic derivations}
In the previous result we have described sufficient and necessary conditions on a $n$-covariant derivation $\delta$ to be continuous on $\bb{K}_\infty$. Besides, we have discussed how a Fréchet continuous derivation $\delta: \bb{K}_\infty \rightarrow \bb{K}_\infty$ can be decomposed as a Fourier series of $n$-covariant derivation. In the next result, which is a restatement of \cite[Theorem~4.5]{Klimek}, we will describe generic continuous derivations on this Fr\'echet algebra as formal commutators.

\begin{theorem}\label{Theorem: Continuous derivations}
	Let $\delta \colon \bb{K}_\infty \to \bb{K}_\infty$ be a continuous derivation.
Then there exists a family of coefficients $\{\beta_{n,l}\}_{n,l \in \Z}$ such that the derivation $\delta$ can be written as the formal commutator
\begin{equation}\label{eq0: Theorem Continuous derivation}
\delta(A) \; = \; \left[\sum_{n,l \in \Z} \beta_{n,l}\, Q_{l,\,l-n},\, A\right],
\end{equation}
and the coefficients $\beta_{n,l}$ satisfy the following growth conditions: for every $\alpha,\gamma \in \N_0$, there exist constants
$$
\tau_i \;\equiv\; \tau_i(\alpha,\gamma) \in \N_0\,, \qquad C_i \;\equiv\; C_i(\alpha,\gamma) > 0\;, \qquad i=1,2\;,
$$
such that
\begin{equation}\label{eq1: Theorem Continuous derivation}
|\beta_{n,l}| \; \leqslant \; C_1 \, \frac{(1+|l-n|)^{\tau_1}}{(1+|n|)^{\gamma}(1+|l|)^{\alpha}}\;,
\end{equation}
and
\begin{equation}\label{eq2: Theorem Continuous derivation}
|\beta_{n,l}| \; \leqslant \; C_2 \, \frac{(1+|l|)^{\tau_2}}{(1+|n|)^{\gamma} (1 + |l - n|)^\alpha}\;.
\end{equation}
Conversely, if a family $\{\beta_{n,l}\}_{n,l \in \Z}$ satisfies~\eqref{eq1: Theorem Continuous derivation} and~\eqref{eq2: Theorem Continuous derivation}, then the formal commutator~\eqref{eq0: Theorem Continuous derivation} defines a continuous derivation on $\bb{K}_\infty$.
\end{theorem}
\begin{proof}
We begin with the converse direction, following a strategy similar to that used in the proof of Proposition~\ref{Prop: commutator n-covariant}. Let $\{\beta_{n, l}\}_{n, l \in \Z}$ be a family of complex numbers satisfying the growth conditions~\eqref{eq1: Theorem Continuous derivation} and~\eqref{eq2: Theorem Continuous derivation}. For any $A \in \bb{K}_\infty$, the formal commutator in Eq.~\eqref{eq0: Theorem Continuous derivation} can be expressed as
\begin{equation}\label{eq4: Theorem Continuous derivation}
\sum_{l,s \in \Z} \left(\sum_{n \in \Z} \beta_{n,l}\, A_{l-n,s}\right) Q_{l,s}
\;-\;
\sum_{l',s' \in \Z} \left(\sum_{n' \in \Z} \beta_{n',\, n' + s'}\, A_{l',\, n' + s'}\right) Q_{l',s'}\;.
\end{equation}
Set
$$ X_{l,s} \; := \;\sum_{n \in \Z} \beta_{n,l} \, A_{l-n,s} \; , \qquad Z_{l',s'} \; := \; \sum_{n' \in \Z} \beta_{n', \, n' + s'} \, A_{l', \, n' + s'} \; .$$

\noindent To show that the formal expression in Eq.~\eqref{eq4: Theorem Continuous derivation} is well defined and yields an element of $\bb{K}_\infty$, we apply Lemma~\ref{Lemma: matrix coefficients} to the families $\{X_{l,s}\}_{l, s \in \Z}$ and $\{Z_{l,s}\}_{l, s \in \Z}$.\\

\noindent Fix $\mu \in \N_0$. We begin with the family $\{X_{l,s}\}_{l, s \in \Z}$. Using the growth condition~\eqref{eq1: Theorem Continuous derivation} with $\alpha = \mu$ and $\gamma = 2$, together with the inequalities $(1 + a + b)^k \leq (1 + a)^k (1 + b)^k$ and $(1 + a)^k (1 + b)^k \leq (1 + a + b)^{2k}$, valid for $a, b \geq 0$ and $k \in \N_0$, we obtain
\begin{equation}\label{eq: Theorem Continuous derivation estimation X}
	\begin{aligned}
		(1 + |l| + |s|)^\mu |X_{l,s}| \; &\leqslant \; (1 + |l| + |s|)^\mu \sum_{n \in \Z} |\beta_{n,l}| \ |A_{l - n, s}| \\
		&\leqslant \; C_{\mu, 2} \, \sum_{n \in \Z} \frac{(1 + |s|)^\mu  (1 + |l|)^{\mu} (1 + |l - n|)^\tau}{(1 + |n|)^2 (1 + |l|)^\mu} |A_{l - n, s}| \\
		&\leqslant \; D_{\mu, 2} \, \sum_{n \in \Z} \frac{(1 + |s|)^{\mu} (1 + |l - n|)^\tau}{(1 + |n|)^2 (1 + |l - n|)^{\tau + \mu} (1 + |s|)^{\tau + \mu}} ||A||_{ 2(\tau + \mu)} \\
		&\leqslant \; M_{\mu, 2} \, ||A||_{ 2(\tau + \mu)} \; .
	\end{aligned} 
\end{equation}
The constant $M_{\mu, 2}$ depends only on $\mu$, and the same holds for $\tau$.

 \smallskip

\noindent For the family $\{Z_{l,s}\}_{l, s \in \Z}$, we proceed similarly, this time using the second growth condition~\eqref{eq2: Theorem Continuous derivation} with $\alpha = \mu$ and $\gamma = 2$. This gives
\begin{equation}\label{eq: Theorem Continuous derivation estimation Z}
	\begin{aligned}
	(1 + |l| + |s|)^\mu |Z_{l,s}| \; &\leqslant \; (1 + |l| + |s|)^\mu \sum_{n \in \Z} |\beta_{n, n + s}| \ |A_{l, n + s}| \\
	&\leqslant \; C_{\mu, 2}' \, \sum_{n \in \Z} \frac{(1 + |s|)^\mu  (1 + |l|)^{\mu} (1 + |n + s|)^\tau}{(1 + |n|)^2 (1 + |s|)^\mu} |A_{l, n + s}| \\
	&\leqslant \; D_{\mu, 2}' \, \sum_{n \in \Z} \frac{(1 + |l|)^{\mu} (1 + |s + n|)^\tau}{(1 + |n|)^2 (1 + |l|)^{\tau + \mu} (1 + |n + s|)^{\tau + \mu}} ||A||_{ 2(\tau + \mu)} \\
	&\leqslant \; M_{\mu, 2}' \, ||A||_{ 2(\tau + \mu)}\;.
	\end{aligned} 
\end{equation}
As before, $M_{\mu,2}$ and $\tau$ only depend on $\mu$. 

 \smallskip

\noindent Since $\mu$ was arbitrary, both families satisfy the decay condition of Lemma~\ref{Lemma: matrix coefficients}. Consequently, the two series in Eq.~\eqref{eq4: Theorem Continuous derivation} converge absolutely and define elements of $\bb{K}_\infty$. Thus, the formal commutator in Eq.~\eqref{eq0: Theorem Continuous derivation} provides a well-defined linear map $\delta_\beta: \bb{K}_\infty \rightarrow \bb{K}_\infty$. Moreover, combining estimates~\eqref{eq: Theorem Continuous derivation estimation X} and~\eqref{eq: Theorem Continuous derivation estimation Z}, we obtain
$$||\delta_\beta(A)||_\mu \; \leqslant \; (M_{\mu, 2} + M'_{\mu, 2}) ||A||_{ 2(\tau + \mu)} \; ,$$
which establishes the Fréchet continuity of $\delta_\beta$.

 \smallskip

\noindent It remains to verify the Leibniz rule. We first observe that the estimates above yield a stronger convergence property. Indeed, fixing $\nu \in \N_0$ and applying estimate~\eqref{eq: Theorem Continuous derivation estimation X} with $\mu = \nu + 4$, we obtain
\begin{align*}
	\sum_{l, s, n \in \Z} ||\beta_{n, l} A_{l - n, s} Q_{l, s}||_\nu \; &= \; \sum_{l, s, n \in \Z} (1 + |l| + |s|)^\nu |\beta_{n, l}|  |A_{l - n, s}| \\
	&\leqslant \; M_{\nu + 4, 2} ||A||_{2(\tau + \nu + 4)} \sum_{l, s} \frac{1}{(1 + |l| + |s|)^4} < \infty \; .
\end{align*}
An analogous estimate, using~\eqref{eq: Theorem Continuous derivation estimation Z}, establishes the same convergence property for the second series in Eq.~\eqref{eq4: Theorem Continuous derivation}. Thus, both series converge absolutely with respect to every defining norm, allowing us to interchange the order of summation.

\noindent In particular, we may perform the summation over $n$ last. For each fixed $n \in \Z$, the corresponding formal commutator defines a continuous derivation by Proposition~\ref{Prop: commutator n-covariant}. The absolute convergence established above, together with the continuity of multiplication, then ensures that $\delta_\beta$ also satisfies the Leibniz rule.

 \smallskip

We now prove the other direction. Let $\delta: \bb{K}_\infty \rightarrow \bb{K}_\infty$ be a Fréchet continuous derivation. Recall that $\delta$ admits a Fourier expansion in terms of its $n$-covariant components, 
$$\delta(A) \; = \; \sum_{n \in \Z} \delta_n(A), \qquad A \in \bb{K}_\infty \; ,$$ 
where the series converges absolutely for each norm $\|\cdot\|_\alpha$.

\noindent By Proposition~\ref{Prop: commutator n-covariant}, for each $n \in \Z$, there exists a sequence $\{\beta_{n,l}\}_{l \in \Z}$ such that $\delta_n$ admits a formal commutator representation of the form~\eqref{eq0: Prop commutator n-covariant}. Combining these representations formally yields the expression~\eqref{eq0: Theorem Continuous derivation}. To justify this representation on $\bb{K}_\infty$, it is enough to establish that the family $\{\beta_{n,l}\}_{n, l \in \Z}$ satisfies the growth conditions~\eqref{eq1: Theorem Continuous derivation} and~\eqref{eq2: Theorem Continuous derivation}. Indeed, once these conditions are verified, the converse direction proved above guarantees that the corresponding formal commutator defines a Fréchet-continuous map on $\bb{K}_\infty$ which coincides with $\delta$.

 \smallskip

\noindent We first show condition~\eqref{eq1: Theorem Continuous derivation}, which requires a separate analysis of the cases $n = 0$ and $n \neq 0$. 

\noindent Let us start with $n = 0$. Applying $\delta_0$ to the matrix unit $Q_{0,l}$ yields $\delta_0(Q_{0,l})=(\beta_{0,0} - \beta_{0,l}) Q_{0,l}$.
Hence,
\begin{align*}
|\beta_{0,l}| \;&\leqslant\; |\beta_{0,l} - \beta_{0,0}| + |\beta_{0,0}| \\
&= \; ||\delta_0(Q_{0,l})||_{0} + |\beta_{0,0}|\\
&\leqslant\; C \, ||Q_{0,l}||_\tau + |\beta_{0,0}| \\
&\leqslant\; (C + |\beta_{0,0}|) (1 + |l|)^\tau\;,
\end{align*}
where $C > 0$ and $\tau \in \N_0$ arise from the continuity of $\delta_0$. 

 \smallskip

\noindent For $n \neq 0$, applying $\delta_n$ to $Q_{l - n,0}$ yields $\delta_{n} (Q_{l - n, 0}) = \beta_{n,l} Q_{l,0} - \beta_{n,0} Q_{l - n, -n}$.
	Thus, for any $\alpha \in \N_0$,
	$$||\delta_n(Q_{l-n,0})||_{\alpha} \; = \; \max \{(1 + |l|)^\alpha |\beta_{n,l}|, (1 + |l - n| + |n|)^\alpha |\beta_{n,0}|\}\;.$$
	Combining this identity with Proposition~\ref{Lemma: decaying Fourier coefficients}, we obtain, for every $\alpha,\gamma \in \N_0$,
	\begin{align*}
		|n|^\gamma (1 + |l|)^\alpha |\beta_{n,l}| \; &\leqslant \; |n|^\gamma ||\delta_n(Q_{l-n,0})||_{\alpha} \\
		&\leqslant \; C' \ ||Q_{l-n,0}||_{\tau'} \\
		&= \; C' \ (1 + |l - n|)^{\tau'}\;, 
	\end{align*}
where $C' > 0$ and $\tau' \in \N_0$ depend only on $\alpha$ and $\gamma$. Since $1 + |n| \leq 2|n|$ for $n \neq 0$, the preceding estimate, together with the one obtained for $n = 0$, proves the growth condition~\eqref{eq1: Theorem Continuous derivation}.
 
  \smallskip
 
\noindent We now turn to condition~\eqref{eq2: Theorem Continuous derivation}. The case $n = 0$ is already covered by the preceding estimate, so we may assume $n \neq 0$. Evaluating $\delta_n$ at $Q_{0, l}$, we find $\delta_n(Q_{0,l}) = \beta_{n,n} Q_{n,l} - \beta_{n,l} Q_{0,l - n}$. Therefore, for any $\alpha \in \N_0$,
	$$||\delta_n(Q_{0,l})||_{\alpha} \; = \; \max \{(1 + |n| + |l|)^{\alpha} |\beta_{n,n}|, (1 + |l - n|)^\alpha |\beta_{n,l}|\}\;.$$
	Using once again Proposition~\ref{Lemma: decaying Fourier coefficients}, we obtain, for any $\alpha,\gamma \in \N_0$,
	\begin{align*}
		|n|^\gamma (1 + |l - n|)^\alpha |\beta_{n,l}| \; &\leqslant \; |n|^\gamma ||\delta_n(Q_{0,l})||_{ \alpha} \\
		&\leqslant \; C'' \ ||Q_{0,l}||_{ \tau''} \\
		&= \; C'' \ (1 + |l|)^{\tau''}\;, 
	\end{align*}
where $C'' > 0$ and $\tau'' \in \N_0$ depend only on $\alpha$ and $\gamma$. This proves condition~\eqref{eq2: Theorem Continuous derivation} and completes the proof. 

\end{proof}

 \smallskip

The previous theorem characterizes all continuous derivations on the Fr\'echet algebra $\bb{K}_\infty$ as commutators with a formal matrix. More precisely, given a continuous derivation $\delta$ with associated coefficients $\{\beta_{n,l}\}_{n,l \in \Z}$, we associate to $\delta$ the formal infinite matrix
\begin{equation}\label{eq: Bdelta}
\mathsf{B}^\delta \; := \; \sum_{n,l \in \Z} \beta_{n,l} \, Q_{l,l-n} \;= \; \sum_{l,r \in \Z} \beta_{l-r,l} \, Q_{l,r}\;.
\end{equation}
Accordingly, the coefficients of $\mathsf{B}^\delta$ are given by $\mathsf{B}^\delta_{i,j} = \beta_{i - j, i}$. 
In this notation, the growth conditions appearing in Theorem~\ref{Theorem: Continuous derivations} can be reformulated as decay estimates on the coefficients of $\mathsf{B}^\delta_{i,j}$. More precisely, for every $\alpha, \gamma \in \N_0$, there exist constants $C > 0$ and $\tau \in \N_0$ such that
\begin{equation}\label{eq: Growth Cond Bdelta}
\Big| \mathsf{B}^\delta_{i,j} \Big|\; \leqslant\;  C \frac{(1 + |j|)^\tau}{(1 + |i - j|)^\gamma (1 + |i|)^\alpha}\;,
\end{equation}
together with the analogous estimate obtained by exchanging the roles of $i$ and $j$ on the left-hand side. 

 \smallskip

In order to further clarify the operator-theoretic content of this description, we now study the action of the associated formal matrix on the Hilbert space $\s{H}$. To this end, we introduce a natural dense subspace that will play an important role in what follows.

\begin{definition}\label{Def: smooth HS}
We define $\s{H}_\infty \subset \s{H}$ as the subspace of vectors $\varphi \in \s{H}$ such that the sequence of coefficients $\big(\langle \psi_k, \varphi \rangle\big)_{k \in \Z}$ is rapidly decaying, that is, for every $\alpha \in \N_0$,
$$
\sup_{k \in \Z} (1 + |k|)^\alpha \, |\langle \psi_k, \varphi \rangle| < \infty.
$$
\end{definition}

 \smallskip

\noindent Equipped with the family of norms
$$
p_\alpha(\varphi) \; := \; \sup_{k \in \Z} (1 + |k|)^\alpha |\langle \psi_k, \varphi \rangle|,
\qquad \alpha \in \N_0\;,
$$
the space $\s{H}_\infty$ becomes a Fr\'echet space that can be regarded as the analogue of a Schwartz space relative to the basis $\Psi$. Moreover, $\s{H}_\infty$ is dense in $\s{H}$, and it coincides with the completion of $\s{H}_{0}^\Psi$ with respect to this family of norms.

\begin{remark}\label{Remark: continuity almost local}
	Every $A \in \bb{K}_\infty$ leaves $\s{H}_\infty$ invariant and defines a Fréchet continuous linear operator on this space. Indeed, for any $\varphi \in \s{H}_\infty$, the action of $A$ can be expressed as 
	\begin{equation*}
		A(\varphi) \; = \; \sum_{k, l \in \Z} \langle \psi_k, \varphi \rangle  \, A_{l, k} \psi_l \; , 
	\end{equation*}
	where, recall, $A_{l, k} = \langle \psi_l, A (\psi_k) \rangle$. To establish that this expression is well defined and converges in the Fréchet topology of $\s{H}_\infty$, we shall see that this double series converges absolutely with respect to all the norms $p_\alpha$. To do that, fix $\alpha \in \N_0$. Since $A \in \bb{K}_\infty$, its matrix coefficients satisfy
	$$|A_{l, k}| \leq \frac{||A||_{2 \alpha + 4}}{(1 + |k| + |l|)^{2\alpha + 4}} \leq \frac{||A||_{2 \alpha + 4}}{(1 + |k|)^{\alpha + 2} (1 + |l|)^{\alpha + 2}} \; .$$
	Consequently, 
	\begin{align*}
	p_\alpha(A (\varphi)) \; &\leqslant \; \sum_{k, l \in \Z} p_\alpha (\langle \psi_k, \varphi \rangle A_{l, k} \psi_l) \\
	&= \; \sum_{k, l \in \Z} |\langle \psi_k, \varphi \rangle| \, |A_{l, k}| \, (1 + |l|)^\alpha \\
	&\leqslant \; ||A||_{2 \alpha + 4} \left(\sum_{l \in \Z} \frac{1}{(1 + |l|)^2}\right) \sum_{k \in \Z} \frac{|\langle \psi_k, \varphi \rangle|}{(1 + |k|)^{\alpha + 2}} \\
	&\leqslant \; ||A||_{2 \alpha + 4} \left(\sum_{l \in \Z} \frac{1}{(1 + |l|)^2}\right) \left(\sum_{k \in \Z} \frac{1}{(1 + |k|)^{2\alpha + 2}}\right) p_\alpha (\varphi) \; .
	\end{align*}
	Since $\alpha$ was arbitrary, the series converges absolutely in the Fréchet topology of $\s H_\infty$. In particular, $A$ leaves $\s H_\infty$ invariant, and the same estimate establishes the Fréchet continuity of its restriction to this space.
	\hfill $\blacktriangleleft$
\end{remark}

For the following proposition, we adopt the following notation. Let $B$ be a possibly unbounded operator on $\s{H}$ such that $\s{H}_\infty \subseteq D(B) \cap D(B^*)$. We denote by
$$ B^\dagger \; := \; B^*|_{\s{H}_\infty}
$$
the restriction of $B^*$ to $\s{H}_\infty$.

\begin{proposition}\label{Proposition: Domain Bdelta}
	Let $\delta : \bb{K}_\infty \to \bb{K}_\infty$ be a Fréchet continuous derivation, and let $\mathsf{B}^\delta$ be the formal matrix associated with $\delta$ as in equation~\eqref{eq: Bdelta}. Then $\mathsf{B}^\delta$ defines a densely defined linear operator $B^\delta$ on $\s{H}$, with domain $D(B^\delta) = \s{H}_\infty$, and which leaves $\s{H}_\infty$ invariant. Moreover, $\s{H}_\infty$ is contained in the domain of its adjoint $(B^\delta)^*$ and is also invariant under this operator. Both	
	$$B^\delta: \s{H}_\infty \rightarrow \s{H}_\infty, \qquad (B^\delta)^\dagger: \s{H}_\infty \rightarrow \s{H}_\infty$$
	are continuous with respect to the Fréchet topology of $\s{H}_\infty$. 
	 
	  \smallskip
	 
	 Conversely, let $(B, D(B))$ be a possibly unbounded operator on $\s{H}$ satisfying $\s{H}_\infty \subseteq D(B)\cap D(B^*)$, and suppose that both $B$ and $B^*$ leave $\s{H}_\infty$ invariant. If their restrictions $B|_{\s{H}_\infty}$ and $B^\dagger$ are Fréchet continuous, then the commutator 
	 \begin{equation*}
	 	\delta_B(A) \; := \; [B, A], \qquad A \in \bb{K}_\infty
	 \end{equation*}
	defines a Fréchet continuous derivation on $\bb{K}_\infty$.
	
	 \smallskip
	
	Finally, if the coefficients of $\mathsf{B}^\delta$ satisfy the anti-hermitian condition 
	\begin{equation}\label{eq: antiHermitian}
	\mathsf{B}^\delta_{j,i} \; = \; -\overline{\mathsf{B}^\delta_{i,j}}\,,
\end{equation}
	then $\ii B^\delta$ is a symmetric operator on the dense domain $\s{H}_\infty$. In particular, 
	$$\delta(A) \; = \; \ii [-\ii B^\delta, A], \qquad A \in \bb{K}_\infty \; .$$
	
\end{proposition}
\begin{proof}
We first establish the properties of the operator associated with a Fréchet continuous derivation. Let $\delta: \bb{K}_\infty \rightarrow \bb{K}_\infty$ be such a derivation, and let $\mathsf{B}^\delta$ be its associated formal matrix, as defined in Eq.~\eqref{eq: Bdelta}. We begin by showing that $\mathsf{B}^\delta$ defines a linear operator on $\s{H}$ with domain $\s{H}_\infty$, leaving this subspace invariant. 

 \smallskip

\noindent Let $\varphi \in \s{H}_\infty$. For each $l \in \Z$, set
\begin{equation*}
	(B^\delta(\varphi))_l \; = \; \sum_{k \in \Z} \langle \psi_k, \varphi \rangle \mathsf{B}^\delta_{l,k} \; .
\end{equation*}
We first show that these coefficients $\{(B^\delta(\varphi))_l\}_{l \in \Z}$ are rapidly decaying. Fix $\alpha \in \N_0$. By the growth condition~\eqref{eq: Growth Cond Bdelta}, there exist constants $C_\alpha > 0$ and $\tau(\alpha) \in \N_0$ such that
$$
|\mathsf{B}^\delta_{l,k}| \; \leqslant\; C_\alpha \frac{(1+|k|)^{\tau(\alpha)}}{(1+|l|)^\alpha}\;.
$$
Using the definition of the norms $p_\mu$, we obtain
\begin{equation}\label{eq2: Prop: derivations in terms of operators}
\begin{aligned}
(1 + |l|)^\alpha |(B^\delta \varphi)_l| \; &\leqslant \; (1 + |l|)^\alpha \sum_{k \in \Z} |\langle \psi_k, \varphi \rangle| \, |\mathsf{B}^\delta_{l,k}| \\
&\leqslant \; C_\alpha \sum_{k \in \Z} (1 + |k|)^{\tau(\alpha)}|\langle \psi_k, \varphi \rangle| \\
&\leqslant \; C_\alpha p_{\tau(\alpha) + 2}(\varphi) \sum_{k \in \Z} \frac{1}{(1 + |k|)^2} \; < \; \infty \; .
\end{aligned}
\end{equation}
Since $\alpha$ was arbitrary, the sequence $\{(B^\delta(\varphi))_l\}_{l \in \Z}$ is rapidly decaying. Hence,
\begin{equation}\label{eq10: Prop: derivations in terms of operators}
	B^\delta (\varphi) \; := \; \sum_{l \in \Z} (B^\delta(\varphi))_l \psi_l 
\end{equation}
defines an element of $\s{H}_\infty$. In fact, an analogous estimate to~\eqref{eq2: Prop: derivations in terms of operators} with a higher index, gives absolute convergence of the corresponding double series in every norm $p_\alpha$. Thus, $B^\delta$ is a densely defined linear operator on $\s{H}$, with domain $D(B^\delta) = \s{H}_\infty$, which, in addition, leaves $\s{H}_\infty$ invariant. Taking the supremum over $l$ in Eq.~\eqref{eq2: Prop: derivations in terms of operators} yields
\begin{equation*}
p_\alpha (B^\delta(\varphi)) \; \leqslant \;  C_\alpha \left(\sum_{k \in \Z} \frac{1}{(1 + |k|)^2}\right) p_{\tau(\alpha) + 2}(\varphi) \; ,
\end{equation*}
which establishes the Fréchet continuity of $B^\delta$ on $\s{H}_\infty$. \\

\noindent We next consider the adjoint of $B^\delta$. For $\xi \in \s{H}_\infty$, define the formal series
\begin{equation}\label{eq4: Prop: derivations in terms of operators}
	\eta \; := \; \sum_{l \in \Z} \left(\sum_{k \in \Z} \langle \psi_k, \xi \rangle \overline{\mathsf{B}^\delta_{k, l}}\right) \psi_l \; .
\end{equation}
We first show that this series defines an element of $\s{H}_\infty$. Fix $\alpha \in \N_0$. By the growth estimate obtained from~\eqref{eq: Growth Cond Bdelta} by exchanging the role of the indices at the left-hand side, there exist $C_\alpha' > 0$ and $\tau'(\alpha) \in \N_0$ such that
$$
|\mathsf{B}^\delta_{k,l}| \; \leqslant\; C'_\alpha \frac{(1+|k|)^{\tau'(\alpha)}}{(1+|l|)^\alpha}\;.
$$
Consequently,
\begin{equation}\label{eq5: Prop: derivations in terms of operators}
	\begin{aligned}
		(1 + |l|)^\alpha \left(\sum_{k \in \Z} \left| \langle \psi_k, \xi \rangle \overline{\mathsf{B}^\delta_{k, l}}\right|\right) \; &\leqslant \; C'_\alpha \sum_{k \in \Z} |\langle \psi_k, \xi \rangle| \, (1 + |k|)^{\tau'(\alpha)} \\
	& \leqslant \; C'_\alpha \sum_{k \in \Z} \frac{1}{(1 + |k|)^2} \, p_{\tau'(\alpha) + 2}(\xi) \; < \; \infty \; .
	\end{aligned}
\end{equation}
It then follows that the series in Eq.~\eqref{eq4: Prop: derivations in terms of operators} defines an element $\eta \in \s{H}_\infty$. Observe that, as for $B^\delta$, estimation~\eqref{eq5: Prop: derivations in terms of operators} with a higher index implies that the double series in the definition of $\eta$ converges absolutely with respect to the family of norms $p_\alpha$.

\noindent To identify $\eta$ with the action of the adjoint, take $\varphi \in \s{H}_\infty$. Then, a formal computation yields
$$\langle \xi, B^\delta (\varphi) \rangle \; = \; \sum_{m, k \in \Z} \overline{\langle \psi_m, \xi \rangle} \langle \psi_k, \varphi \rangle \mathsf{B}_{m, k}^\delta \; = \; \langle \eta, \varphi \rangle \; .$$
The rapid decay of the coefficients of $\xi$ and $\varphi$, together with the growth bounds on $\mathsf{B}^\delta$, ensures absolute convergence of the double series and justifies the interchange of summations. It then follows that $\xi \in D((B^\delta)^*)$ and that $(B^\delta)^\dagger(\xi) = \eta$. Thus, one has
$$\s{H}_\infty \subseteq D((B^\delta)^*), \qquad (B^\delta)^\dagger (\s{H}_\infty) \subseteq \s{H}_\infty \; .$$
Finally, taking the supremum over $l$ in~\eqref{eq5: Prop: derivations in terms of operators} gives
$$p_\alpha ((B^\delta)^\dagger (\xi)) \; \leqslant \; C_\alpha' \left(\sum_{k \in \Z}\frac{1}{(1 + |k|)^2}\right) p_{\tau'(\alpha) + 2}(\xi) \; ,$$
which proves the Fréchet continuity of $(B^\delta)^\dagger$ on $\s{H}_\infty$. \\

We now turn to the converse. Let $(B, D(B))$ be a densely defined linear operator on $\s{H}$ such that $\s{H}_\infty \subseteq D(B) \cap D(B^*)$, and such that both $B|_{\s{H}_\infty}$ and $B^\dagger$ are Fréchet-continuous operators on $\s{H}_\infty$. Consider the family of coefficients
\begin{equation*}
	\mathsf{B}_{k, l} \; := \; \langle \psi_k, B (\psi_l) \rangle, \qquad k,l \in \Z \; .
\end{equation*}
We show that the family $\{\mathsf{B}_{k, l}\}_{k, l \in \Z}$ fulfils the growth condition~\eqref{eq: Growth Cond Bdelta} as well as the one obtained by interchanging the roles of $k$ and $l$.

\noindent Fix $\alpha, \gamma \in \N_0$. Using the Fréchet continuity of $B|_{\s{H}_\infty}$ and the inequality $1 + |k - l| \leq (1 + |k|) (1 + |l|)$, we obtain
\begin{equation}\label{eq7: Prop: derivations in terms of operators}
	\begin{aligned}
		|\mathsf{B}_{k, l}| \; &\leqslant \; \frac{p_{\alpha + \gamma}(B (\psi_l))}{(1 + |k|)^{\alpha + \gamma}} \\
		&\leqslant \; C_{\alpha + \gamma} \frac{p_{\tau(\alpha + \gamma)}(\psi_l)}{(1 + |k|)^{\alpha + \gamma}} \\
		&\leqslant \; C_{\alpha + \gamma} \frac{(1 + |l|)^{\tau(\alpha + \gamma) + \gamma}}{(1 + |k|)^\alpha (1 + |k - l|)^\gamma} \; .
	\end{aligned} 
\end{equation}

\noindent The second growth condition follows similarly from the Fréchet continuity of $B^\dagger$. Indeed,
\begin{equation}\label{eq8: Prop: derivations in terms of operators}
	\begin{aligned}
		|\mathsf{B}_{k, l}| \; = \; |\langle \psi_l, B^\dagger (\psi_k) \rangle| \; &\leqslant \; \frac{p_{\alpha + \gamma}(B^\dagger (\psi_k))}{(1 + |l|)^{\alpha + \gamma}} \\
		&\leqslant \; C'_{\alpha + \gamma} \frac{p_{\tau'(\alpha + \gamma)}(\psi_k)}{(1 + |l|)^{\alpha + \gamma}} \\
		&\leqslant \; C'_{\alpha + \gamma} \frac{(1 + |k|)^{\tau'(\alpha + \gamma) + \gamma}}{(1 + |l|)^\alpha (1 + |k - l|)^\gamma} \; .
	\end{aligned} 
\end{equation}
According to Theorem~\ref{Theorem: Continuous derivations}, the coefficients $\mathsf{B}_{k, l}$ define a Fréchet continuous derivation $\delta_\mathsf{B}: \bb{K}_\infty \rightarrow \bb{K}_\infty$. Moreover, the first part of the proof shows that $\mathsf{B}$ defines a Fréchet continuous operator $\widehat{B}: \s{H}_\infty \rightarrow \s{H}_\infty$ via Eq.~\eqref{eq10: Prop: derivations in terms of operators}. By construction, $\widehat{B}$ agrees with $B|_{\s{H}_\infty}$ on $\s{H}_0^\Psi$, and hence on all of $\s{H}_\infty$ by continuity and density. 

\noindent Now, by Remark~\ref{Remark: continuity almost local}, every $A \in \bb{K}_\infty$ leaves $\s{H}_\infty$ invariant and defines a Fréchet continuous operator on this space. Consequently, the commutator $[B,A]$ is well defined on $\s{H}_\infty$. Moreover, for every $i, j \in \Z$,
\begin{equation*}
	\begin{aligned}
	\langle \psi_i, (BA - AB)(\psi_j)\rangle \; = \; \sum_{k \in \Z} \bigl(\mathsf{B}_{i, k} A_{k, j} - A_{i, k} \mathsf{B}_{k, j}\bigr) \; .
\end{aligned}
\end{equation*}
These coefficients define an element in $\bb{K}_\infty$ by Lemma~\ref{Lemma: matrix coefficients}. Indeed, using that $A \in \bb{K}_\infty$ and the growth condition~\eqref{eq7: Prop: derivations in terms of operators} yields, for a fixed $\alpha \in \N_0$,
\begin{align*}
(1 + |i| + |j|)^\alpha \sum_{k \in \Z} |\mathsf{B}_{i, k}| \, |A_{k, j}| \; &\leqslant \; C_\alpha \sum_{k \in \Z} (1 + |j|)^\alpha (1 + |k|)^{\tau(\alpha)} \, |A_{k, j}| \\
&\leqslant \; C_{\alpha} ||A||_{\beta} \sum_{k \in \Z} \frac{1}{(1 + |k|)^2}, 
\end{align*}
with $C_\alpha > 0$ and $\beta = \max \{\alpha, \tau(\alpha)\} + 2$. A similar computation works for $\sum_{k \in \Z} A_{i, k} \mathsf{B}_{k, j}$ using the growth condition~\eqref{eq8: Prop: derivations in terms of operators}. Notice, in particular, that these estimations, together those ones in the proof of Lemma~\ref{Lemma: matrix coefficients}, show that the triple series $\sum_{i, j \in \Z}\big(\sum_{k \in \Z} (\mathsf{B}_{i, k} A_{k, j} - A_{i, k} \mathsf{B}_{k, j})\big) Q_{i, j}$ converges absolutely with respect to the norms $\|\cdot\|_\alpha$.

\noindent Therefore, $[B, A] \in \bb{K}_\infty$ and, clearly, it satisfies the Leibniz identity. Furthermore, from this computation in matrix units, one concludes that $[B, A] = \delta_\mathsf{B}(A)$.\\

Finally, by following a similar argument, one shows that if the matrix coefficients satisfy the anti-Hermitian condition~\eqref{eq: antiHermitian}, then the associated operator $\ii B^\delta$ is symmetric on $\s{H}_\infty$.
\end{proof}

\section{$0$--Chains}\label{Section: 0-chains}

Having characterized the continuous derivations on the Fréchet algebra $\bb{K}_\infty$, we now introduce the notion of a $0$-chain. This description is motivated by quantum many-body systems, where interactions naturally give rise to derivations expressed in terms of spatially distributed contributions (see, for instance,~\cite{Kapustin-Sopenko-Local-Noether}). From a mathematical perspective, the formulation in terms of $0$-chains provides a convenient framework for studying derivations through the spatial growth of their defining interactions, with a view toward extending this approach to more general algebras.

\noindent In the present setting of compact operators, our main result, Proposition~\ref{Prop: Growth-Condition 0 chain}, establishes a sufficient growth condition for a $0$-chain to define a Fréchet-continuous derivation. Unlike the results of the preceding section, this does not provide a complete characterization of continuous derivations on $\bb{K}_\infty$. Rather, its significance lies in the form of the condition itself, which is formulated in terms of spatial growth and serves as a starting point for the more general framework to be developed in subsequent work.

\begin{definition}\label{Def: 0-Chains}
Let $\Phi: \Z \to \bb{K}_\infty$ be a function such that
$ \Phi(k)^* = \Phi(k)$
 for every $k \in \Z$,
and such that 
$ |||\Phi(k)|||_{k,\alpha} < \infty$ for every $\alpha \in \N_0$.
We refer to functions of this form as \emph{$0$-chains}. Such a function $\Phi$ defines, formally, a derivation $\delta_\Phi$ as 
\begin{equation}\label{eq: 0-chain Derivation}
	\delta_{\Phi}(A) \; := \; \ii \, \sum_{k \in \Z} [\Phi(k), A].
\end{equation}
The collection of all $0$--chains $\Phi$ is denoted by $C_0(\bb{K}_\infty)$.
\end{definition}

\begin{remark}\label{Rem: 0-chain literature}
The notion of $0$--chain interaction adopted here slightly differs from the one commonly used in the literature. In the context of quantum spin systems, it is assumed in the definition of these maps the stronger uniform bound
$$
\sup_{k \in \Z} \, |||\Phi(k)|||_{k,\alpha} \; < \; \infty \qquad \forall\;\alpha \in \N_0\; .
$$
In the present work, we deliberately relax this uniformity condition. This weaker notion is sufficient for our purposes and allows us to encompass a larger class of continuous derivations on the Fr\'echet algebra $\bb{K}_\infty$.
\hfill $\blacktriangleleft$
\end{remark}

\subsection{$0$--Chains and Fr\'echet continuous derivations}\label{Subsection: 0-chains and derivations}
Our goal is to determine when the formal commutator~\eqref{eq: 0-chain Derivation} defines a continuous derivation on the Fr\'echet algebra $\bb{K}_\infty$. The next proposition provides a sufficient condition ensuring a positive answer. In addition, we describe how to associate a $0$--chain to any given symmetric continuous derivation $\delta$.
In the following we will use  $\Phi(k)_{i,j}$ for the $(i,j)$-matrix coefficient of $\Phi(k)$ with respect to the basis $\Psi$.

\begin{proposition}\label{Prop: interaction and derivation}
Let $\Phi$ be a $0$-chain. Then the formal derivation $\delta_{\Phi}$ defined in~\eqref{eq: 0-chain Derivation} defines a Fréchet continuous derivation on the Fr\'echet algebra $\bb{K}_\infty$ provided that, for every $\alpha, \gamma \in \N_0$, there exist constants $C > 0$ and $\tau \in \N_0$ such that
\begin{equation}\label{eq: 0chain-growth}
\sum_{k \in \Z} \big|\Phi(k)_{i,j}\big|
\;\leqslant\;
C\, \frac{(1 + |i|)^\tau}{(1 + |j|)^\alpha (1 + |i - j|)^\gamma},
\end{equation}
for all $i, j \in \Z$.  \\

Conversely, let $\delta : \bb{K}_\infty \to \bb{K}_\infty$ be a Fréchet continuous symmetric derivation with associated anti-Hermitian matrix $\mathsf{B}^\delta$. Then $\delta$ induces a $0$-chain map $\Phi$ satisfying condition~\eqref{eq: 0chain-growth}.
\end{proposition}

\begin{proof}
Let $\Phi$ be a $0$-chain satisfying condition~\eqref{eq: 0chain-growth}. According to Theorem~\ref{Theorem: Continuous derivations}, it suffices to construct a family of matrix coefficients $\mathsf{B}^{\delta_{\Phi}}_{i,j}$ satisfying the growth conditions stated there. To this end, set
	$$(\mathsf{B}^{\delta_{\Phi}})_{i,j} \; := \; \ii \sum_{k \in \Z} \Phi(k)_{i,j} \; .	$$
By assumption~\eqref{eq: 0chain-growth}, the series converges absolutely for every $i,j \in \Z$. Since each $\Phi(k)$ is self-adjoint, $\sum_{k \in \Z} |\Phi(k)_{i, j}| = \sum_{k \in \Z} |\Phi(k)_{j, i}|$. Hence, it also follows the required growth conditions.

\noindent Hence, $\mathsf{B}^{\delta_{\Phi}}$ defines a Fréchet continuous derivation $\delta_{\Phi}$ on $\bb{K}_\infty$, given explicitly by the formal commutator
$$ \delta_{\Phi}(A) \; = \; \ii \left[ \sum_{k \in \Z} \Phi(k), A \right] \; .$$

For the converse, let $\delta: \bb{K}_\infty \rightarrow \bb{K}_\infty$ be a Fréchet continuous symmetric derivation with associated anti-Hermitian matrix $\mathsf{B}^{\delta}$. We define a $0$-chain $\Phi : \Z \to \bb{K}_\infty$ using matrix coefficients as follows. Set
	\begin{equation}\label{eq: From derivation to 0-chain}
		\Phi(k)_{m, n} \; = \; \begin{cases}
		-\ii \, (\mathsf{B}^{\delta})_{m,k}, &\text{if } n = k \text{  and  } m \geq k\\[4pt]
	-\ii \, (\mathsf{B}^{\delta})_{k, n}, &\text{if } m = k \text{   and  } n \geq k + 1 \\[4pt]
	0, & \text{otherwise}\;.
		\end{cases}
	\end{equation}	
Notice that, by the anti-Hermiticity of $\mathsf{B}^\delta$, $\Phi(k)_{m, n} = \overline{\Phi(k)_{n, m}}$. In order to see that these matrix coefficients define an element of $\bb{K}_\infty$, we apply Lemma~\ref{Lemma: matrix coefficients}. Fix $\alpha \in \N_0$. Using the anti-hermitian property of $\mathsf{B}^\delta$ and its growth estimates, we obtain
\begin{align*}
	\sup_{m, n \in \Z}(1 + |m| + |n|)^\alpha |\Phi(k)_{m, n}| \; &= \; \sup_{m \geq k} (1 + |k| + |m|)^\alpha \left|\mathsf{B}^\delta_{m, k}\right| \\
	&\leqslant \; C_\alpha (1 + |k|)^{\alpha + \tau(\alpha)} \; ,
\end{align*}
where $C_\alpha > 0$ and $\tau(\alpha) \in \N_0$ are constants arising from the growth condition~\eqref{eq: Growth Cond Bdelta}. Thus, for each $k \in \Z$, the resulting operator $\Phi(k)$ belongs to $\bb{K}_\infty$. Moreover, from its construction, $\Phi(k)^* = \Phi(k)$ and
	\begin{equation}\label{eq: From derivation to interaction spatial estimate}
	\|\Phi(k)\|_{k, \alpha} \; \leqslant \; (1 + 2|k|)^\alpha\,\|\Phi(k)\|_{0, \alpha} \; \leqslant \; \widetilde{C}_\alpha \, (1 + |k|)^{2\alpha + \tau(\alpha)} \; ,
\end{equation}
	It remains to verify condition~\eqref{eq: 0chain-growth}. This follows by observing that the $0$-chain $\Phi$ corresponds to a foliation of the matrix $\mathsf{B}^\delta$. Hence, one gets that $\sum_{k \in \Z} |\Phi(k)_{i,j}| = |\mathsf{B}^\delta_{i,j}|$.
\end{proof}

\subsection{Spatial growth conditions for Fréchet continuity}

The description of derivations in terms of $0$-chains developed in the previous Subsection suggests that their continuity may be studied through the spatial growth of the individual terms $\Phi(k)$. For instance, in the context of quantum spin systems over a lattice $\Gamma$, it is known that uniform $0$-chains, satisfying bounds of the form
$$
\sup_{k\in\Gamma}|||\Phi(k)|||_{k,\alpha} \; < \; \infty,
$$
for every $\alpha\in\N_0$, induce continuous derivations on the corresponding Fréchet algebra of almost-local observables~\cite{Kapustin-Sopenko-Local-Noether}. In the present setting of compact operators, however, such uniform bounds need not hold, even when the associated derivation is continuous.

\noindent Such construction developed above shows that polynomial growth can always be achieved by choosing a suitable representative of the same derivation. More precisely, by the norm equivalence in Proposition~\ref{Proposition: Equivalence norms compact algebra} and estimate~\eqref{eq: From derivation to interaction spatial estimate}, if a $0$-chain $\Phi$ induces a continuous symmetric derivation $\delta_\Phi$, then there exists another $0$-chain $\Phi'$ such that $\delta_{\Phi'} = \delta_\Phi$ and, for every $\alpha \in \N_0$, there are constants $C_\alpha > 0$ and $\theta(\alpha) \in \N_0$ satisfying
$$ |||\Phi'(k)|||_{k, \alpha} \; \leqslant \; C_\alpha (1 + |k|)^{\theta(\alpha)}, \qquad k \in \Z \; .$$

\noindent This suggests that suitable restrictions on the spatial growth of $0$-chains may provide sufficient conditions for continuity, even in the absence of uniform bounds. In the following proposition, we establish such a condition for the compact-operator algebra $\bb{K}_\infty$. Although this result does not provide a complete characterization of continuous derivations, as obtained in the preceding sections, its formulation in terms of spatial growth will be particularly useful for the more general framework that we intend to develop in subsequent work.

\begin{proposition}\label{Prop: Growth-Condition 0 chain}
	Let $\Phi : \Z \to \bb{K}_\infty$ be a $0$--chain map satisfying the following power--type control of the spatial seminorms: for every $\alpha \in \N_0$ there exist constants $C_\alpha > 0$ and $s_\alpha \in \R_{\geq 0}$ such that
\begin{equation}\label{eq: power-type-growth}
|||\Phi(k)|||_{k,\alpha} \; \leqslant\; C_\alpha \, (1 + |k|)^{s_\alpha}, \qquad k \in \Z.
\end{equation}
Assume in addition that the exponents $\{s_\alpha\}_{\alpha \in \N_0}$ satisfy
\begin{equation}\label{eq: limSup powerBounds}
	\limsup_{\alpha \to \infty} (\alpha - s_\alpha)  \; = \; +\infty .
\end{equation}
Then the derivation $\delta_{\Phi}$ associated with the $0$-chain $\Phi$ is continuous with respect to the Fr\'echet topology on $\bb{K}_\infty$.
\end{proposition}

\begin{proof}
	For each $k \in \Z$, we use a telescoping decomposition of an element $B \in \bb{K}_\infty$ into operators localized around $k$. Set $B_{k, 0} := 0$ and, for $r \in \N$, define
$$ B_{k, r} \; := \; P_{k, r} B P_{k, r} \; . $$
Writing $B^{(r)}_k := B_{k, r + 1} - B_{k, r}$ for $r \in \N_0$, we have, for every $\beta \in \N_0$,

\begin{equation}\label{eq0: Bounding telescoping terms}
\|B^{(r)}_k\|  \; \leqslant \; \|B - B_{k, r + 1}\| + \|B - B_{k, r}\|  \; \leqslant \; \frac{4}{(1 + r)^\beta} \, |||B|||_{k, \beta}.
\end{equation}
Since $B_{k, r} \to B$ in norm as $r \to \infty$, this gives the telescoping expansion
$$ B = \sum_{r \in \N_0} B^{(r)}_k \; , $$
which, in addition, by estimate~\eqref{eq0: Bounding telescoping terms}, converges absolutely with respect to every norm $|||\cdot|||_{k, \beta}$.

\noindent Fix $A \in \bb{K}_\infty$. Applying this decomposition to each $\Phi(k)$, with localization centered at $k$, and to $A$, with localization centered at $0$, we obtain 
$$ \Phi(k) \; = \; \sum_{s = 0}^\infty \Phi(k)^{(s)}_k, \qquad \Phi(k)^{(s)}_k \in \bb{K}_{k, s + 1} \; ,$$
and
$$ A \; = \; \sum_{r = 0}^\infty A^{(r)}_0, \qquad A^{(r)}_0 \in \bb{K}_{0, r + 1} \; . $$

\noindent Formally, the associated derivation $\delta_\Phi$ can therefore be expressed as
$$ \delta_\Phi(A)  \; = \; \ii \sum_{k \in \Z} \sum_{r = 0}^\infty \sum_{s = 0}^\infty [\Phi(k)^{(s)}_k, A^{(r)}_0].$$
Fix $\alpha \in \N_0$. By locality, the commutator $[\Phi(k)^{(s)}_k, A^{(r)}_0]$ vanishes whenever $|k| > r + s + 2$. For $|k| \leqslant r + s + 2$, the inclusion $\Phi(k)^{(s)}_k \in \bb{K}_{0, 3 + 2s + r}$ implies that
$$[\Phi(k)^{(s)}_k, A^{(r)}_0] \in \bb{K}_{0, 3 + 2s + r} \; .$$ 
Using the definition of the norms $|||\cdot|||_{0, \alpha}$ and estimate~\eqref{eq0: Bounding telescoping terms}, we obtain
\begin{equation}\label{eq1: estimate Prop continuity 0 chains}
	\begin{aligned}
		|||[\Phi(k)^{(s)}_k, A^{(r)}_0]|||_{0, \alpha} \; &\leqslant \; 4 (4 + 2s + r)^\alpha ||\Phi(k)^{(s)}_k|| \, ||A^{(r)}_0|| \\
	&\leqslant \; D_\alpha (1 + s + r)^\alpha \frac{|||\Phi(k)|||_{k, \beta}}{(1 + s)^\beta} \frac{|||A|||_{0, \gamma}}{(1 + r)^\gamma},
	\end{aligned}
\end{equation}
where $D_\alpha > 0$ depends only on $\alpha$.  \\

\noindent We also have the following estimate. Using the growth condition \eqref{eq: power-type-growth}, we obtain
\begin{equation}\label{eq2: estimate Prop continuity 0 chains}
	\begin{aligned}
	\sum_{\substack{k \in \Z \\ |k| \leqslant r + s + 2}} |||\Phi(k)|||_{k, \beta} \; \leqslant \; C_\beta \sum_{\substack{k \in \Z \\ |k| \leqslant r + s + 2}} (1 + |k|)^{s_\beta} \; \leqslant \; C'_\beta (1 + r + s)^{s_\beta + 1}
	\end{aligned}
\end{equation}
where $C'_\beta = 2 \cdot 3^{s_\beta + 1} C_\beta$ and, in the second inequality, we have used that the set $\{-(r + s + 2), \ldots, r + s + 2\}$ contains $2(r+s) + 5$ integers. Combining~\eqref{eq1: estimate Prop continuity 0 chains} and~\eqref{eq2: estimate Prop continuity 0 chains}, we obtain
	\begin{align}\label{eq3: estimate Prop continuity 0 chains}
		|||\delta_\Phi(A)|||_{0, \alpha} \; &\leqslant \; \sum_{r \geq 0} \sum_{s \geq 0} \sum_{\substack{k \in \Z \\ |k| \leqslant r + s + 2}} |||[\Phi(k)^{(s)}_k, A^{(r)}_0]|||_{0, \alpha}  \notag \\
	&\leqslant \; D_\alpha |||A|||_{0, \gamma} \sum_{r \geq 0} \sum_{s \geq 0} \frac{(1 + s + r)^{\alpha}}{(1 + s)^\beta (1 + r)^\gamma} \sum_{\substack{k \in \Z \\ |k| \leqslant r + s + 2}} |||\Phi(k)|||_{k, \beta}  \notag \\
	&\leqslant \; D_\alpha  C'_\beta  |||A|||_{0, \gamma} \sum_{r \geq 0} \sum_{s \geq 0}\frac{(1 + s + r)^{\alpha + s_\beta + 1}}{(1 + s)^\beta (1 + r)^\gamma} \notag \\
	&\leqslant \;  D_\alpha C'_\beta  |||A|||_{0, \gamma} \left(\sum_{s \geq 0}\frac{(1 + s)^{\alpha + s_\beta + 1}}{(1 + s)^\beta}\right)\left(\sum_{r \geq 0}\frac{(1 + r)^{\alpha + s_\beta + 1}}{(1 + r)^\gamma}\right) \; .
	\end{align}
By condition \eqref{eq: limSup powerBounds}, we can choose $\beta$ sufficiently large such that the sum over $s$ above converges. Once $\beta$ is fixed, we then choose $\gamma$ sufficiently large to ensure convergence of the series over $r$. Thus, the triple series converges absolutely in each norm $|||\cdot|||_{0,\alpha}$. 

\noindent Since the telescoping expansions of $\Phi(k)$ and $A$ converge absolutely in the Fréchet sense, for each $k \in \Z$ we have
$$ [\Phi(k), A] \; = \; \sum_{r, s \geq 0} [\Phi(k)_k^{(s)}, A_0^{(r)}] \; .$$
The absolute convergence established in~\eqref{eq3: estimate Prop continuity 0 chains} allows us to rearrange the sums, showing that the triple series coincides with the formal expression $\ii \sum_{k\in\Z}[\Phi(k),A]$. Hence, $\delta_\Phi(A)$ is well defined, and this estimate also establishes the Fr\'echet continuity of $\delta_\Phi$.
\end{proof}

\section{Dynamics and derivations}\label{Section: Non-integrable derivations}

In the previous section, we reformulated the study of continuous symmetric derivations into the framework of $0$-chains, establishing sufficient conditions for a $0$-chain to induce such a derivation. We now demonstrate that continuity alone is insufficient to guarantee the existence of a dynamics. In this section, we provide an explicit example of a continuous derivation $\delta: \bb{K}_\infty \rightarrow \bb{K}_\infty$ that fails to integrate into a strongly continuous one-parameter group of automorphisms on the Fr\'echet algebra $\bb{K}_\infty$.

\subsection{Integrable continuous derivations}

We begin by making precise the notion of a derivation generating a one-parameter group.

\begin{definition}\label{Definition: One Parameter Frechet}
	Let $\delta: \bb{K}_\infty \rightarrow \bb{K}_\infty$ be a Fr\'echet continuous derivation. We say that $\delta$ generates a strongly continuous one-parameter group of automorphisms $T_\delta: \R \rightarrow \operatorname{Aut}(\bb{K}_\infty)$ (throughout, $\operatorname{Aut}$ denotes the group of $\ast$-automorphisms) if the following conditions hold:
	\begin{enumerate}
		\item For each $t \in \R$, the map $T_\delta(t): \bb{K}_\infty \rightarrow \bb{K}_\infty$ is continuous. That is for any $\alpha \in \N_0$, there exist $C > 0$ and $\beta \in \N_0$, depending on $t$ and $\alpha$, such that
		$$||T_\delta(t)(A)||_{ \alpha}\; \leqslant\; C \ ||A||_{\beta}, \qquad \forall\, A \in \bb{K}_\infty\;.$$
		\item $T_\delta$ satisfies the group property, namely
		\[
		T_\delta(0)\; =\; \operatorname{Id}_{\bb{K}_\infty}\;,\qquad
		T_\delta(t + s)\; =\; T_\delta(t) \circ T_\delta (s)
		\]  for all $t,s \in \R$.
		\item For each $A \in \bb{K}_\infty$, the map $t \mapsto T_\delta(t)(A)$ is continuous.
		\item For each $A \in \bb{K}_\infty$ and $t_0 \in \R$, 
		$$\lim_{h \to 0} \frac{T_\delta(t_0 + h)(A) - T_{\delta}(t_0)(A)}{h} \; = \; T_\delta(t_0)(\delta(A)),$$
		where the limit is taken in the Fr\'echet sense. 
	\end{enumerate}
\end{definition}

With this definition in place, a natural question is whether such a dynamical system extends to the entire algebra $\bb{K}$ of compact operators. We address this in the following proposition.
\begin{proposition}\label{Prop: extension to K}
	Let $\delta: \bb{K}_\infty \rightarrow \bb{K}_\infty$ be a Fr\'echet continuous derivation, and suppose that $\delta$ integrates to a strongly continuous one-parameter group $T_\delta: \R \rightarrow \operatorname{Aut}(\bb{K}_\infty)$. Then, $T_\delta$ extends to a strongly continuous one parameter group (with respect to the operator norm) of automorphisms $\widetilde{T}: \R \rightarrow \operatorname{Aut}(\bb{K})$.
\end{proposition}
\begin{proof}
Fix $t \in \R$, and let $r > 0$. Since $\bb{K}_{0,r} \subset \bb{K}_{\rm loc} \subset \bb{K}_\infty$, by restricting $T_\delta$ to $\bb{K}_{0,r}$ one gets a $*$-homomorphism from $\bb{K}_{0,r}$ to $\bb{K}$. Hence, for any $A \in \bb{K}_{0,r}$, one has $||T_\delta(t)(A)|| \leqslant ||A||$. It then follows that $T_\delta(t)$ restricts to a norm contracting $*$-homomorphism over $\bb{K}_{\rm loc}$, and by density, one gets a unique extension $\widetilde{T}(t): \bb{K} \rightarrow \bb{K}$. Since one can do the same analysis for $T_\delta(-t)$, it follows that, in fact, $\widetilde{T}(t)$ is an automorphism on the algebra of compact operators with inverse $\widetilde{T}(-t)$. Taking $A \in \bb{K}_\infty$, by density with respect to the Fr\'echet topology, there exists a sequence of finite supported elements $\{A_n\}_{n \in \N} \subset \bb{K}_{\rm loc}$ such that $A_n$ converges to $A$ in the Fr\'echet sense. Since $||\cdot|| \leqslant S_2 ||\cdot||_{2}$ by Eq.~\eqref{remark: Operator norm vs matrix norm}, this sequence also converges to $A$ in the norm topology. Hence, one gets
$$\widetilde{T}(t)(A) \;=\; \lim_{n \to \infty} \widetilde{T}(t)(A_n) \;=\; \lim_{n \to \infty} T_\delta(t)(A_n) = T_\delta(t)(A).$$ 
Thus, $\widetilde{T}(t)|_{\bb{K}_\infty} = T_\delta(t)$.
Now, let $t, s \in \R$, and let $A \in \bb{K}_\infty$. One has that $\widetilde{T}(t + s)(A) = T_\delta(t + s)(A) = T_\delta(t)\big(T_\delta(s)(A)\big),$ but $T_\delta(s)(A) \in \bb{K}_\infty$, and we have already observed that $\widetilde{T}(t)$ restricted to $\bb{K}_\infty$ is equal to $T_\delta(t)$, so $T_\delta(t) (T_\delta(s)(A)) = \widetilde{T}(t)(T_\delta(s)(A)) = \widetilde{T}(t)(\widetilde{T}(s)(A))$. Since $\bb{K}_\infty$ is dense, the same holds for any $A \in \bb{K}$.
It remains to show that $\widetilde{T}$ is strongly continuous. To show that, fix $t_0$ and $A \in \bb{K}_\infty$. Using again the equivalence of norms, one gets
\begin{align*}
	\lim_{t \to t_0} ||\widetilde{T}(t)(A) - \widetilde{T}(t_0)(A)|| \; & = \; \lim_{t \to t_0} ||T_\delta(t)(A) - T_\delta(t_0)(A)||\\
	& \leqslant \; C \lim_{t \to t_0}||T_\delta(t)(A) - T_\delta(t_0)(A)||_{ 2} = 0\;.
\end{align*}
Let now $A \in \bb{K}$. Let $\varepsilon > 0$, and let $B \in \bb{K}_\infty$ such that $||A - B|| < \varepsilon/3$. Since $T_\delta$ is strongly continuous, there exists $\eta > 0$ such that for any $t \in \R$ with $|t - t_0| < \eta$ one has 
$$||\widetilde{T}(t)(B) - \widetilde{T}(t_0)(B)|| \;\leqslant\; C \ ||T_\delta(t)(B) - T_\delta(t_0)(B)||_{2} \;<\; \frac{\varepsilon}{3}.$$
Thus, one obtains:
\begin{align*}
	||\widetilde{T}(t)(A) - \widetilde{T}(t_0)(A)||\;  \leqslant\;  &||\widetilde{T}(t)(A - B)|| + ||\widetilde{T}(t)(B) - \widetilde{T}(t_0)(B)||\\ &+ ||\widetilde{T}(t_0)(B) - \widetilde{T}(t_0)(A)||\\
	\leqslant\;& 2 \ ||A - B|| + ||\widetilde{T}(t)(B) - \widetilde{T}(t_0)(B)|| \;<\; \varepsilon\;.
\end{align*}
Thus, $\widetilde{T}: \R \rightarrow \operatorname{Aut}(\bb{K})$ defines a strongly continuous, with respect to the operator norm, one-parameter group of automorphisms over the algebra of compact operators such that, restricted to the Fr\'echet algebra $\bb{K}_\infty$, coincides with $T_\delta$.
\end{proof}

 \smallskip

Having extended the dynamics to the $C^*$-algebra $\bb{K}$, we now describe its infinitesimal generator in the usual sense of $C^*$-dynamical systems. Let $\widetilde{T} : \R \to \mathrm{Aut}(\bb{K})$ be the strongly continuous one-parameter group obtained above. Its infinitesimal generator is defined by
$$
\Delta(A) \; := \; \lim_{t\to 0} \frac{\widetilde{T}(t)(A)-A}{t},
$$
whenever the limit exists in the operator norm. The domain of the generator is
$$
D(\Delta) \; := \; \left\{ A \in \bb{K}\;\middle|\; \lim_{t\to 0} \frac{\widetilde{T}(t)(A)-A}{t}\;\;
\text{exists in norm} \right\}\;.
$$
It is well known that $\Delta$ is a densely defined closed operator on $\bb{K}$, where closedness is understood with respect to the operator norm topology, i.e. its graph is closed in $\bb{K} \oplus \bb{K}$ endowed with the norm topology (see \cite[\S IX]{Yosida}).

 \smallskip

\noindent Since $\widetilde{T}(t)$ coincides with $T_\delta(t)$ on $\bb{K}_\infty$, it follows from Definition~\ref{Definition: One Parameter Frechet} that for every $A \in \bb{K}_\infty$ the limit
$$
\lim_{t\to 0} \frac{\widetilde{T}(t)(A)-A}{t}\; =\; \lim_{t\to 0} \frac{T_\delta(t)(A)-A}{t}\; =\; \delta(A)
$$
exists in the Fr\'echet topology, and therefore also in the operator norm. Hence,
$$
\bb{K}_\infty\; \subseteq\; D(\Delta)\;, \qquad \Delta|_{\bb{K}_\infty} \;=\; \delta\;.
$$
Moreover, the norm-closure of $\delta$, viewed as a densely defined operator on $\bb{K}$, coincides with $\Delta$. Indeed, since $\bb{K}_\infty$ is norm-dense in $\bb{K}$ and invariant under the one-parameter group $\widetilde{T}$, it follows from~\cite[Corollary 3.1.7]{Bratteli-Robinson-1} that $\bb{K}_\infty$ is a core for $\Delta$. Consequently, $\overline{\delta} = \Delta$.

 \smallskip

Since we are dealing with the $C^*$-algebra of compact operators on a separable Hilbert space, the extended one-parameter group $\widetilde{T} : \mathbb{R} \longrightarrow \mathrm{Aut}(\bb{K})$ admits a concrete implementation.
According to \cite[Theorem 1, 4]{articleBratteli}, there exists a (possibly unbounded) self-adjoint operator $H$ with dense domain $D(H) \subseteq \s{H}$ such that for all $t \in \R$ and $A \in \bb{K}$,
$$\widetilde{T}(t)(A) \;=\; e^{\ii tH} A e^{-\ii tH} \; .$$
It then follows that $\Delta(\cdot) = \ii [H, \cdot]$, with the additional property that for any $A \in D(\Delta)$, $A D(H) \subseteq D(H)$ (see \cite[Proposition 3.2.55]{Bratteli-Robinson-1}). Moreover, this implementing operator $H$ is unique modulo additive scalar multiples of the identity, that is, $H + \lambda \operatorname{Id}_{\s{H}}$ generates the same one-parameter group for any $\lambda \in \mathbb{R}$.

 \smallskip

By Proposition~\ref{Proposition: Domain Bdelta}, every symmetric Fréchet-continuous derivation $\delta$ gives rise to a symmetric operator $-\ii B^\delta$, defined on the dense subspace $\s{H}_\infty$ and continuous with respect to its Fréchet topology. We shall show that the closure of $-\ii B^\delta$ in the Hilbert space $\s{H}$ coincides with $H$, up to an additive scalar multiple of the identity.

\begin{proposition}\label{Prop: Essentially self adjointness}
Let $\delta : \bb{K}_\infty \to \bb{K}_\infty$ be a continuous symmetric derivation, and let $-\ii B^\delta : \s{H}_\infty \subseteq \s{H} \to \s{H}$ be the associated symmetric operator constructed in Proposition~\ref{Proposition: Domain Bdelta}. Suppose that $\delta$ generates a strongly continuous one-parameter group $T_\delta$, and let $\widetilde{T}$ be its extension to $\bb{K}$ with infinitesimal generator $\Delta$ with domain $D(\Delta) \subseteq \bb{K}$ implemented by a self-adjoint operator $H$ on $\s{H}$. Then the operator $-\ii B^\delta$ is closable in $\s{H}$ and its closure coincides with $H$ up to an additive scalar multiple of the identity, that is,
$$
\overline{-\ii\,B^\delta} \; = \; H + \lambda \operatorname{Id}_{\s{H}}
$$
for some $\lambda \in \mathbb{R}$.
\end{proposition}

\begin{proof}
	Fix a unit vector $\xi \in \s{H}_\infty$, and for each $\varphi \in \s{H}_\infty$ consider the rank-one operator $P_{\varphi, \xi} (\ \cdot\ ) = \langle \xi, \ \cdot \rangle \varphi$. Since $\s{H}_\infty$ consists of rapidly decaying vectors, $P_{\varphi,\xi}$ belongs to $\bb{K}_\infty$.\\
	
\noindent Because $\Delta$ coincides with $\delta$ on $\bb{K}_\infty$, one has
	\begin{equation}\label{eq0: Proof Essentially selfadjointess}
		\ii [-\ii B^\delta, P_{\varphi, \xi}] \; = \; \ii [H, P_{\varphi, \xi}].
	\end{equation}
	By~\cite[Proposition 3.2.55]{Bratteli-Robinson-1}, for any $\chi \in D(H)$ and $A \in D(\Delta)$ the commutator $\ii [H,A](\chi)$ is well defined and given by
	$$\ii \ H(A (\chi)) -\ii \ A(H(\chi)).$$
	Using the density of $D(H)$, choose $\chi \in D(H)$ such that $\langle\xi,\chi\rangle \neq 0$. Taking $A  = P_{\varphi,\xi}$, it follows that $\varphi \in D(H)$ and
	$$H(\varphi) = \frac{1}{\langle \xi, \chi \rangle} \left([H,P_{\varphi, \xi}](\chi) + \langle \xi, H (\chi)\rangle \varphi\right).$$
	 Hence, $\s{H}_\infty \subseteq D(H)$. Applying both sides of \eqref{eq0: Proof Essentially selfadjointess} to $\xi$ yields
	 $$-\ii B^\delta(\varphi) \; = \; H(\varphi) + (\langle \xi, -\ii B^\delta (\xi)\rangle - \langle \xi, H(\xi) \rangle) \varphi.$$
	 Thus $B^\delta$ and $H$ coincide on $\s{H}_\infty$ up to an additive scalar multiple of the identity, which without loss of generality may be assumed to be zero. 
To conclude, observe that the one-parameter group of unitary operators $e^{\ii tH}$ leaves $\s{H}_\infty$ invariant. Indeed, using the operator $P_{\varphi, \xi}$ and the dynamics $\widetilde{T}$, one obtains
	 $$P_{e^{\ii tH}(\varphi), e^{\ii tH}(\xi)} = \widetilde{T}(t)(P_{\varphi, \xi}) = T_\delta(t)(P_{\varphi, \xi}) \in \bb{K}_\infty, \qquad \forall t \in \R,$$
	 and therefore $e^{\ii tH}(\varphi) \in \s{H}_\infty$. Applying~\cite[Corollary 3.1.7]{Bratteli-Robinson-1} to the one-parameter group $e^{\ii tH}$, it follows that $\s{H}_\infty$ is a core for $H$, and consequently $\overline{-\ii B^\delta} = H$.
\end{proof}

\begin{corollary}\label{Corollary: Necessary condition essential self-adjointness}
Let $\delta: \bb{K}_\infty \to \bb{K}_\infty$ be a Fréchet-continuous symmetric derivation. If $\delta$ generates a strongly continuous one-parameter group of automorphisms on $\bb{K}_\infty$, then the associated symmetric operator $-\ii B^\delta$, with domain $\s{H}_\infty$, is essentially self-adjoint.
\end{corollary}

\subsection{Non-integrable continuous derivations}
One might expect that continuity of a symmetric derivation on $\bb{K}_\infty$ is sufficient to ensure that it generates a strongly continuous one-parameter group of automorphisms. However, this is not the case. We show this by constructing a class of matrices satisfying the growth conditions of Theorem~\ref{Theorem: Continuous derivations}, whose associated symmetric operators fail to be essentially self-adjoint. By Corollary~\ref{Corollary: Necessary condition essential self-adjointness}, the corresponding derivations cannot generate such groups. Our construction is based on Jacobi matrices (see~\cite{teschl} for a comprehensive treatment).

 \smallskip

Let $\eta > 1$ and consider the sequence
$$a_n \; = \; \begin{cases} 
			0 & \text{if } n < 0\;, \\ 
			(1 + n)^\eta & \text{otherwise}\;. 
		\end{cases}$$

\noindent Define the infinite Hermitian matrix
$$
\mathsf{A} \; = \;
\begin{pmatrix}
\ddots & \ddots & \ddots &        &        &        \\
\ddots & 0 & a_{-2} &   0     &        &        \\
  \ddots     & a_{-2} & 0      & a_{-1} &     0   &        \\
       &   0     & a_{-1} & 0      & a_0    &    0    \\
       &        &    0    & a_0    & 0      & a_1   & \ddots  \\
       &        &        &   0     & a_1    & 0      & \ddots \\
       &        &        &        &    \ddots    & \ddots & \ddots
\end{pmatrix}.
$$
Notice that $\mathsf{B} := \ii \mathsf{A}$ is anti-Hermitian in the sense of~\eqref{eq: antiHermitian} and satisfies the growth condition~\eqref{eq: Growth Cond Bdelta}. Indeed, $B_{i,j} = 0$ unless $|i-j|=1$, and
$$|\mathsf{B}_{n, n+1}| \; = \; |\mathsf{B}_{n + 1, n}| \; = \; |a_n| \; \leqslant \; (1 + |n|)^\eta \; .$$
Thus, for every $\alpha, \gamma \in \N_0$,
$$(1 + |n|)^\alpha |\mathsf{B}_{n, n+1}| \; \leqslant \; (1 + |n|)^{\alpha + \eta} \; .$$
Since all other matrix coefficients vanish, these estimates establish the growth condition~\eqref{eq: Growth Cond Bdelta} for all $i,j \in \Z$. According to Theorem~\ref{Theorem: Continuous derivations}, $\mathsf{B}$ defines a symmetric Fréchet continuous derivation on $\bb{K}_\infty$. Moreover, the associated $0$-chain, constructed as
in~\eqref{eq: From derivation to 0-chain}, satisfies
$$|||\Phi(k)|||_{k, \alpha} \; \leqslant \; 2 \|\Phi(k)\| \; \leqslant \; 2 \, (1 + |k|)^\eta\;,
\qquad \alpha \in \N_0\;,$$
so that the spatial growth of $\Phi$ is controlled in the sense of Proposition~\ref{Prop: Growth-Condition 0 chain}.

 \smallskip

Let $A_{\mathrm{min}}$ denote the operator obtained by letting this matrix act on the space $\s{H}_0^\Psi$. Since $a_n = 0$ for $n < 0$, the operator $A_{\mathrm{min}}$ decomposes into the direct sum of the zero operator and a half-line Jacobi operator, the latter with domain $\operatorname{span}\{\psi_n: n \in \N_0\}$. According to~\cite[Corollary 1]{Kostyuchenko}, this Jacobi operator has deficiency indices $(1,1)$. Since the zero operator is self-adjoint, $A_{\mathrm{min}}$ has the same deficiency indices and, in particular, is not essentially self-adjoint.\\

One might attempt to remedy this by extending the operator to the larger domain $\s{H}_\infty$. Let $A$ denote the operator defined by the same matrix on $\s{H}_\infty$. By Proposition~\ref{Proposition: Domain Bdelta}, $A$ is Fr\'echet continuous as an operator $A : \s{H}_\infty \to \s{H}_\infty$. Nevertheless, this extension still does not produce a core for any self-adjoint extension of $A_{\mathrm{min}}$. Indeed, we claim that $\s{H}_\infty \subseteq  D(\overline{A_{\mathrm{min}}})$. Let $\varphi \in \s{H}_\infty$. Since $\s{H}_0^\Psi$ is dense in $\s{H}_\infty$ with respect to the Fr\'echet topology, there exists a sequence $\{\varphi^m\}_{m \in \N} \subset \s{H}_0^\Psi$ converging to $\varphi$ in the Fr\'echet sense. By continuity of $A$, the sequence $\{A(\varphi^m)\}$ converges to $A(\varphi)$ in the Fr\'echet topology. Since $\norm{\varphi} \leqslant c \; p_1(\varphi)$ for every $\varphi \in \s{H}_\infty$, with $c^2 := \sum_{k \in \Z}(1 + |k|)^{-2}$, this convergence also holds in $\s{H}$. By the definition of the closure of $A_{\mathrm{min}}$, it follows that $\varphi \in D(\overline{A_{\mathrm{min}}})$.

 \smallskip

We summarize the previous discussion in the following theorem.

\begin{theorem}\label{Theorem: No lifted derivation}
There exists a Fr\'echet-continuous symmetric derivation $\delta : \bb{K}_\infty \to \bb{K}_\infty$
which does not generate any one-parameter group of automorphisms of $\bb{K}_\infty$ that is strongly
continuous with respect to the Fr\'echet topology, in the sense of Definition~\ref{Definition: One Parameter Frechet}.
\end{theorem}

\newpage

\appendix

\section{Rapidly approximate sequences}\label{app:Rap_seq}
Let $\s{S}(\N_0)$ be the space of complex valued sequences that decay at infinity faster than any power (Schwartz-type sequences), \ie
\[
\s{S}(\N_0)\;:=\;\{g:\N_0\to\C\;|\;||g||_k<\infty\;,\;\; \forall\; k\in\N_0 \}
\]
where the system of norms $||\cdot||_k$ is defined by
 \begin{equation}\label{eq:seq_nor}
||g||_k\;:=\;\sup_{r\in\N_0}\;(1+r)^k|g(r)|\;<\;+\infty\;,\qquad \forall\;k\in\N_0
\end{equation}
Since $(1+r)^{k+1}\geqslant (1+r)^k$
 it follows that $||g||_k\leqslant ||g||_{k+1}$ for every $g\in\s{S}(\N_0)$, meaning that the system of norms is increasing.
\begin{proposition}\label{prop:compl}
 The vector space  $\s{S}(\N_0)$ turns out to be a  a Fr\'echet space when topologized by the family  of norms $||\cdot||_k$. 
\end{proposition}

 \smallskip

Observe that 
\[
(1+r)^k\;|f(r)g(r)|\;\leqslant\;(1+r)^{2k}\;|f(r)|\;|g(r)|
\]
one immediately deduces that 
\begin{equation*}
||fg||_k\;\leqslant\;||f||_k\;||g||_k\;,\qquad \forall\; k\in\N_0\;
\end{equation*}
\ie all the norms are submultiplicative with the pointwise product.
The space $\s{S}(\N_0)$ can also be endowed with the involution given by the complex conjugation. If $g\in\s{S}(\N_0)$ then $g^*$ is the sequence defined by $g^*(r):=\overline{g(r)}$. Clearly $g^*\in \s{S}(\N_0)$ and one can easily check that
\begin{equation*}
||gg^*||_k\;\leqslant\;||g||_k^2\;=\;||g^*||_k^2\;,\qquad \forall\; k\in\N_0\;.
\end{equation*}
Furthermore, any non-negative function $f \in \s{S}(\N_0)$ can be bounded above by a monotonically decreasing function $g \in \s{S}(\N_0)$. For instance, one can define $g$ by:
\begin{equation*}
 g(r) \; := \; \sup_{s \geqslant r} f(s).
\end{equation*}
\begin{proposition}\label{pro:F-alg}
 The space  $\s{S}(\N_0)$ is a Fr\'echet $\ast$-algebra with respect to the pointwise multiplication and the complex conjugation.
\end{proposition}

%======================================================================
\renewcommand{\refname}{References}

% Bibliography inlined from Bibliography.bib, ordered by first appearance
% in the text. Only the entries actually cited are kept; the unused ones
% (Bratteli-Robinson-2, Higson, Kapustin-Sopenko-Lieb, Komorowski,
% Lieb-RobinsonOriginal, Naaijkens-peter, nachtergaele-scholz-werner-13,
% RaeburnMorita, Varilly) have been dropped.
% Unless marked otherwise, the data is transcribed from Bibliography.bib
% and has NOT been independently verified against the sources.

\end{document}